\documentclass[manuscript]{aastex}
\usepackage{emulateapj5}
\usepackage{graphicx}
\usepackage{times}

\slugcomment{Accepted for publication ApJ, 2006 Jun 13}

\shortauthors{A. M. Hopkins, J. F. Beacom}
\shorttitle{On the normalisation of the Cosmic SFH}

\begin{document}

\title{On the normalisation of the cosmic star formation history}

\author{Andrew M. Hopkins\altaffilmark{1}, John F. Beacom\altaffilmark{2,3}
}

\affil{
\begin{enumerate}
\item School of Physics, University of Sydney, NSW 2006, Australia;
email: ahopkins@physics.usyd.edu.au
\item Dept.\ of Physics, The Ohio State University, 191 W.\ Woodruff Ave.,
Columbus, OH 43210; email: beacom@mps.ohio-state.edu
\item Dept.\ of Astronomy, The Ohio State University, 140 W.\ 18th Ave.,
Columbus, OH 43210
\end{enumerate}
}

\begin{abstract}
Strong constraints on the cosmic star formation history (SFH) have recently
been established using ultraviolet and far-infrared measurements, refining
the results of numerous measurements over the past decade. Taken together,
the most recent and robust data indicate a compellingly consistent picture of
the SFH out to redshift $z \approx 6$, with especially tight constraints for
$z \lesssim 1$. We fit these data with simple analytical forms, and derive
conservative bands to indicate possible variations from the best fits.
Since the $z \lesssim 1$ SFH data are quite {\em precise}, we investigate
the sequence of assumptions and corrections that together affect the SFH
normalisation, to test their {\em accuracy}, both in this redshift range and
beyond. As {\em lower} limits on this normalisation, we consider the evolution
in stellar mass density, metal mass density, and supernova rate density,
finding it unlikely that the SFH normalisation is much lower than indicated
by our direct fit. Additionally, predictions from the SFH for supernova
type~Ia rate densities tentatively suggests delay times of $\sim 3$\,Gyr. As
a corresponding {\em upper} limit on the SFH normalisation, we consider the
Super-Kamiokande (SK) limit on the electron antineutrino
($\overline{\nu}_e$) flux from past core-collapse supernovae, which applies
primarily to $z\lesssim 1$. We find consistency with the SFH only if the
neutrino temperatures from SN events are relatively modest. Constraints on the
assumed initial mass function (IMF) also become apparent. The traditional
Salpeter IMF, assumed for convenience by many authors, is known to be a poor
representation at low stellar masses ($\lesssim 1\,M_{\odot}$), and we show
that recently favoured IMFs are also constrained. In particular, somewhat
shallow, or top-heavy, IMFs may be preferred, although they cannot be too
top-heavy. To resolve the outstanding issues, improved data are called for on
the supernova rate density evolution, the ranges of stellar masses leading
to core-collapse and type~Ia supernovae, and the antineutrino and neutrino
backgrounds from core-collapse supernovae.
\end{abstract}

\keywords{galaxies: evolution --- galaxies: formation ---
 galaxies: starburst --- neutrinos --- supernovae: general}

\section{Introduction}
\label{int}

In the past few years measurement of the evolution of galaxy luminosity
functions at a broad range of wavelengths has rapidly matured. One consequence
of this has been the refinement in our understanding of how the space density
of the galaxy star formation rate (SFR) evolves \citep{Lil:96,Mad:96}.
In particular the recent results from the Sloan Digital Sky Survey (SDSS),
the Galaxy Evolution Explorer (GALEX), and Classifying Objects by Medium-Band
Observations (COMBO17) at ultraviolet (UV) wavelengths, and from the Spitzer
Space Telescope at far-infrared (FIR) wavelengths, now allows this cosmic star
formation history (SFH) to be quite tightly constrained (to within
$\approx30-50\%$) up to redshifts of $z\approx1$. Combined with measurements
of the SFH at higher redshifts from FIR, sub-millimeter, Balmer line and
UV emission, the SFH is reasonably well determined (within a factor of about
3 at $z\gtrsim1$) up to $z\approx6$ \citep[e.g.,][]{Hop:04}.

Additional results from the Super-Kamiokande (SK) particle detector provide
a strong limit on the electron antineutrino ($\overline{\nu}_e$) flux,
1.2\,cm$^{-2}$\,s$^{-1}$ (for $E_{\nu}>19.3\,$MeV), originating from supernova
type~II (SNII; here and throughout, we assume the inclusion of all
core-collapse supernovae: types II, Ib, and Ic) events associated with the
SFH \citep{Mal:03}. This limit on the diffuse supernova neutrino background
(DSNB) acts to constrain the normalisation of the SFH
\citep{FK:03,And:04a,Str:04,Str:05}. An exploration of quantities predicted
from the SFH, the stellar and metal mass density evolution, and supernova (SN)
rate evolution, provides further insight into the allowable normalisation of
the SFH \citep{Str:05}. This series of interconnected physical properties of
galaxies and SNe provides an emerging opportunity for determining
the level of the SFH normalisation, and the SFH measurements particularly
for $z\lesssim 1$ now have the precision to allow this exploration of
their accuracy. Constraining the normalisation of the SFH will support
a range of quantitative analyses of galaxy evolution, including the
mass-dependence of the SFH \citep[e.g.,][]{Pap:06, Jun:05, Hea:04}, and the
reasons underlying the decline in the SFH to low redshifts
\citep[e.g.,][]{Bell:05}.

The sequence of assumptions explored here starts from
observed luminosity density measurements. (1)~To these, dust corrections
(where necessary), SFR calibrations and IMF assumptions are applied, to
produce corrected SFH measurements. (2)~From the SFH, assumptions about the
high-mass IMF fraction that produce SNII lead to predictions for the SNII
rate density. This can be compared directly with measurements of this
quantity. (3)~Assumptions about the neutrino emission per SNII then give
a prediction of the DSNB, for comparison with the SK limit. Since optical
SNII can be hidden from observations by dust obscuration, the present SNII
rate density measurements may merely be lower limits. In contrast, since
neutrinos are unaffected by dust, the DSNB provides an absolute upper 
limit on the true SNII rate. The consistency that we find between the SFH
data and the SK upper limit on the DSNB indicates that there is little room
to increase the overall normalisation of the SFH. Thus, this constrains
each of the dust corrections, the assumed IMF, and the neutrino emission per
supernova. None can be increased significantly without requiring at least one
of the others to be reduced below its expected range. Below, we focus
especially on the assumed IMF, noting the more detailed treatment of the
neutrino emission per supernova by \citet{Yuk:05}. A parallel set of
assumptions to step (2) regarding the generation of SNIa lead to predictions
for the SNIa rate density, and this is explored in some detail with
tantalising implications regarding the extent of the SNIa delay time.

In \S\,\ref{data} we update the SFH data compilation of \citet{Hop:04}
and address some of the assumptions that affect the normalisation.
We identify the best parametric fit to the most robust subset of this data
in \S\,\ref{fits}, consistent with the $\overline{\nu}_e$ limits from SK.
In \S\,\ref{results} we present the results of this fitting in terms of
the stellar and metal mass density evolution and the SNII and SNIa rate
evolution. The implications for the assumed IMF and SNIa properties are
discussed further in \S\,\ref{disc}.

The 737\footnote{Thanks to Sandhya Rao \citep{Rao:06} for this terminology.}
cosmology is assumed throughout with $H_0=70\,$km\,s$^{-1}$\,Mpc$^{-1}$,
$\Omega_M=0.3$, $\Omega_{\Lambda}=0.7$ \citep[e.g.,][]{Spe:03}.

\section{The data}
\label{data}
\begin{figure*}
\centerline{\rotatebox{-90}{\includegraphics[width=10cm]{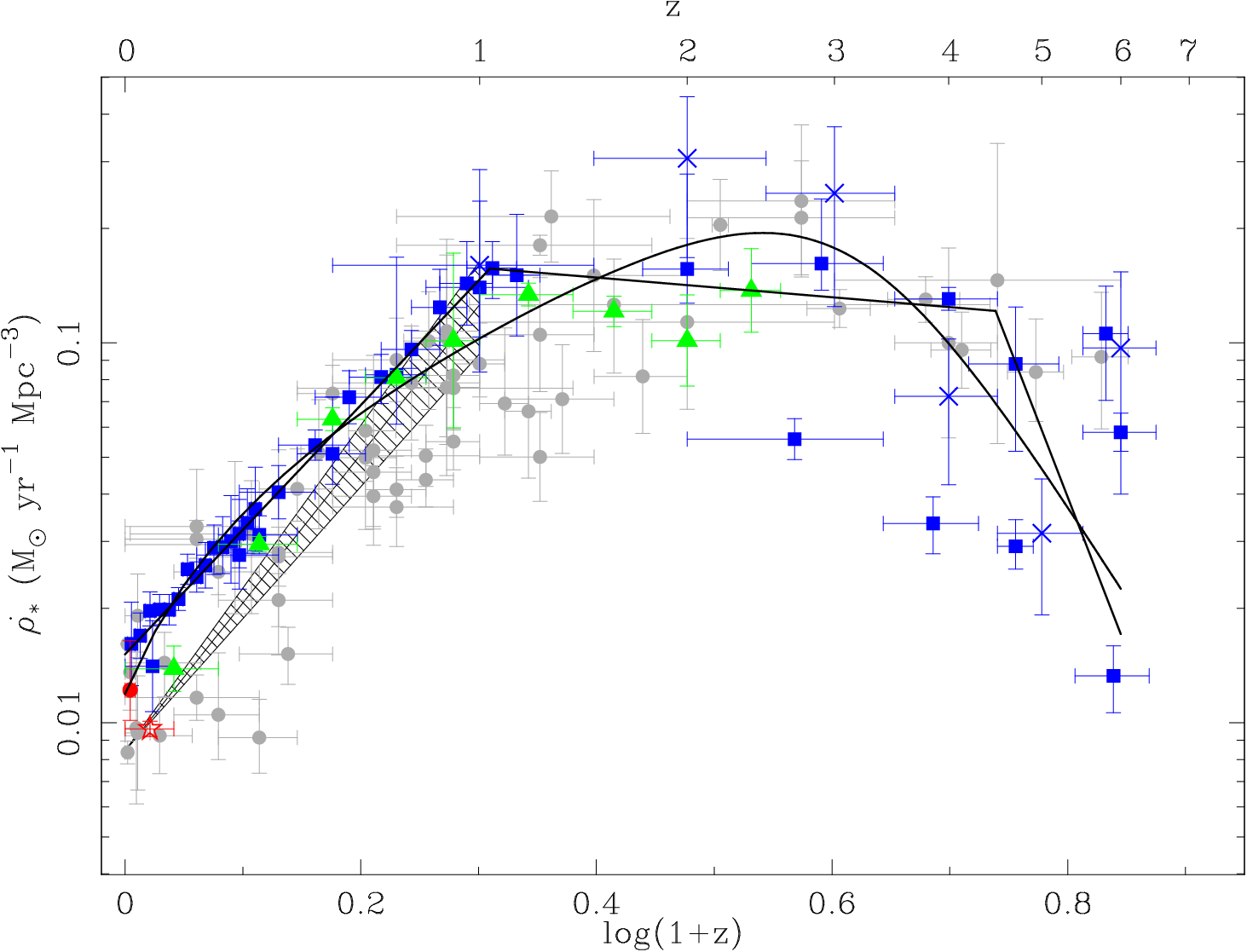}}}
\caption{Evolution of SFR density with redshift. Data shown here have been
scaled assuming the SalA IMF. The grey points are from the compilation
of \citet{Hop:04}. The hatched region is the FIR ($24\,\mu$m) SFH
from \citet{LeF:05}. The green triangles are FIR ($24\,\mu$m) data from
\citet{Per:05}. The open red star at $z=0.05$ is based on radio
(1.4\,GHz) data from \citet{Mau:05}. The filled red circle at $z=0.01$
is the H$\alpha$ estimate from \citet{Han:06}. Blue squares are UV data from
\citet{Bal:05,Wol:03,Arn:05,Bou:03b,Bou:03a,Bun:04,Bou:05,Ouch:04}. Blue
crosses are the UDF estimates from \citet{Tho:06}. Note that these have
been scaled to the SalA IMF assuming they were originally estimated using
a uniform \citet{Sal:55} IMF.
The solid lines are the best-fitting parametric forms (see text for details
of which data are used in the fitting).
Although the FIR SFH of \citet{LeF:05} is not used directly in the fitting,
it has been used to effectively obscuration-correct the UV data to the values
shown, which are used in the fitting. Note that the top logarithmic scale
is labelled with redshift values, not $(1+z)$.
 \label{fig:sfh}}
\end{figure*}

The compilation of \citet{Hop:04} was taken as the starting point for this
analysis, shown in Figure~\ref{fig:sfh} as grey points. These data are
reproduced from Figure~1 of \citet{Hop:04}, and use their ``common"
obscuration correction where necessary. Additional measurements are
indicated in colour in Figure~\ref{fig:sfh}. For $z\lesssim3$ these
consist of FIR ($24\,\mu$m) photometry from the Spitzer Space Telescope
\citep{Per:05,LeF:05}, and UV measurements from the SDSS \citep{Bal:05},
GALEX \citep{Arn:05,Sch:05} and the COMBO17 project \citep{Wol:03}. At
$z=0.05$ a new radio (1.4\,GHz) measurement is shown \citep{Mau:05}, which is
highly consistent with the FIR results, as expected from the radio-FIR
correlation \citep{Bell:03}. Also at low redshift ($z=0.01)$ is a new
H$\alpha$ derived measurement \citep{Han:06}. At higher redshifts additional
SFH measurements come from the Hubble Ultra Deep Field \citep[UDF,][]{Tho:06},
and from various photometric dropout analyses, probing rest-frame
UV luminosities \citep{Bou:03a,Bou:03b,Ouch:04,Bun:04,Bou:05}. The
UDF measurements of \citet{Tho:06} are derived through fitting spectral
energy distributions to the UDF photometry using a variety of templates
with a range of underlying assumptions. In particular this includes
different IMF assumptions for different templates. Although we show these
measurements in Figure~\ref{fig:sfh} for illustrative purposes (having
scaled them assuming they were uniformly estimated using a
\citeauthor{Sal:55} \citeyear{Sal:55} IMF), we do not include them in
subsequent analyses as there is no clear process for scaling these
measurements to our assumed IMFs in the absence of a common original
IMF assumption.

\subsection{SFR calibrations}
\label{sfrcal}
Throughout we assume the same SFR calibrations as \citet{Hop:04}.
Uncertainties in the calibrations for different SFR indicators will correspond
to uncertainties in the resulting SFH normalisation for that indicator. Issues
regarding SFR calibrations are detailed in \citet{Mou:06}, \citet{Ken:98} and
\citet{Con:92}. Perhaps the most uncertain calibrations are the radio
(1.4\,GHz) and FIR indicators (although the [O{\sc ii}] indicator has a
similar level of uncertainty). For FIR SFRs, \citet{Ken:98} indicates a
variation of about 30\% between calibrations in the literature.
\citet{Bell:03} refines the 1.4\,GHz calibration of \citet{Con:92} following
an exploration of the origins of the radio-FIR correlation, and the
implication is that the radio SFR calibration has about the same
uncertainty as the FIR, assuming no contamination by emission from an
active galactic nucleus (AGN). More significantly, though, for individual
galaxies there can be large differences, up to an order of magnitude,
in the SFRs inferred through different indicators \citep[e.g.][]{Hop:03},
although on average for large samples there is a high level of consistency.
This is reflected in the overall consistency between SFR densities,
$\dot{\rho}_*$, estimated from different indicators, with at most about a
factor of two or three variation (which also includes the uncertainty in
dust obscuration corrections, where relevant). This scatter is still notably
larger than the uncertainties in individual SFR calibrations, and is suggestive
of the overall level of systematic uncertainty in the individual calibrations.
It is likely that this reflects subtleties such as low-level AGN contamination
in various samples, the difficulties with aperture corrections where
necessary, dust obscuration uncertainties (discussed further below), and
other issues. It is for these reasons that we neglect the details of the
underlying SFR calibrations, as their small formal uncertainties are dominated
by these larger systematics. Further, the effect of these systematics between
different SFR indicators acts to increase the scatter in the overall SFH
compilation, rather than to systematically shift all measurements in
a common direction. So even the factor of two to three variation here
cannot be viewed precisely as an uncertainty on the SFH normalisation. In
this sense, the level of consistency between $\dot{\rho}_*$ measurements
using dramatically different samples and SFR indicators, over the whole
redshift range up to $z\approx 6$, is actually quite encouraging.

\subsection{Dust obscuration corrections}
\label{dust}
The issue of dust corrections is complex and has been addressed by many
authors \citep[e.g.,][]{Buat:05,Bell:03b,Buat:02,Cal:01}.
\citet{Hop:04} compared assumptions of a ``common" obscuration correction
to a luminosity-dependent correction on the SFH. The latter leads, for UV and
emission line estimators, to somewhat higher values for $\dot{\rho}_*$
at higher redshift, although both methods give results essentially
consistent with SFH estimators unaffected by obscuration. \citet{Bell:03}
shows that SFRs derived from summing total IR ($8-1000\,\mu$m) and UV
indicators using the SFR calibrations of \citet{Ken:98} adopted here, are
consistent within a factor of 2 of obscuration corrected H$\alpha$ SFRs.
This technique, suggested as the preferred method by \citet{Igl:06}, has
the strong advantage that it avoids assumptions about the extent or form
of the obscuration, and variations due to possible luminosity bias in
the UV selected sample \citep[e.g.,][]{Hop:01,Sul:01,Afo:03}.

To implement effective obscuration corrections for the UV measurements at
$z\lesssim 1$ \citep{Bal:05,Wol:03,Arn:05}, we thus take advantage of the
well-established FIR SFR densities up to $z=1$ from \citet{LeF:05}. The UV data
at $z\le1$ are ``obscuration corrected" by adding the FIR SFR density from
\citet{LeF:05} to each point. As shown by \citet{Bell:03} for individual
systems, this technique results in $\dot{\rho}_*$ estimates consistent with
the obscuration corrected H$\alpha$ measurements, in particular the
recent estimates from \citet{Han:06} and \citet{Doh:06},
\citep[also compare the current Figure~\ref{fig:sfh} to Figure~1 of][]{Hop:04}.
This result is consistent with the interpretation of \citet{Tak:05} that
about half the SFR density in the local universe is obscured by dust
\citep[see also][]{Mar:05}, increasing to about 80\% by $z\approx1$,
a trend that can be seen in the different slopes of the (obscuration
corrected) UV measurements and FIR measurements in Figure~\ref{fig:sfh}.
For obscuration corrections to the UV data between $1<z<3$ we rely on
the fact that the FIR measurements of \citet{Per:05} are quite flat in
this domain, as well as being highly consistent with those of \citet{LeF:05}
at $z<1$, and add the constant SFR density corresponding to that
of \citet{LeF:05} at $z=1$. This is also consistent with the recent
measurements of obscuration corrections for UV luminosities at
$z\approx 2$ by \citet{Erb:06}, who find a typical correction factor
of $\approx 4.5$. At higher redshifts we apply a ``common"
obscuration correction to the UV data as detailed in \citet{Hop:04}.
The reliability of this assumption is open to question, but may not be
unreasonable, as a comparison with the the measured colour excesses of
\citet{Ouch:04} illustrates. The obscuration corrected SFR densities
of \citet{Ouch:04} at $z=4-5$ using their measured colour excesses
are actually marginally {\em higher} than we derive with the ``common"
obscuration correction.

\subsection{UV data at high-$z$}
\label{uvhighz}
At $z\gtrsim 2$, some clarification is necessary regarding the UV derived
$\dot{\rho}_*$ measurements. The two UV measurements at $z\approx 2$ and
$z\approx 3$ are taken from the Hubble Deep Field (HDF) sample of
\citet{Arn:05}, and have been corrected for obscuration by adding to them
a constant FIR SFR density equal to that at $z=1$ from \citet{LeF:05}.
The three UV points at $2\lesssim z\lesssim 5$ with comparatively low
$\dot{\rho}_*$ come from \citet{Bou:03a}, and are based on photometric dropouts
using the cloning technique detailed by those authors. These points are low
as they correspond to only the high-$L$ end of the luminosity function,
and are not used here in any subsequent analysis. The two UV measurements
at $z\approx 4$ and $z\approx 5$ are from \citet{Ouch:04}, corrected using
a ``common" obscuration correction, and are marginally lower than the
$\dot{\rho}_*$ derived by those authors using their measured $E(B-V)$
colour excesses. This slight difference has a negligible impact on the
fitted parametric forms detailed below, and does not affect any subsequent
analysis. At $z\approx 6$ there are now a large number of estimates in the
literature, primarily using the photometric dropout technique and hence
relying on accurate photometry and colour measurements, all of which are based
on small, deep fields (including the HDF, the Great Observatories Origins Deep
Survey, GOODS, and the UDF). The two highest
measurements of $\dot{\rho}_*$ at $z\approx 6$ come from GOODS \citep{Gia:04}
and $i$-dropouts in two deep ACS fields \citep{Bou:03b}. These estimates
appear to be high compared to subsequent photometric dropout analyses, and
this seems to be a result of sample contamination due to large colour
uncertainties from low S/N photometric measurements (A.\ Bunker 2005, private
communication). These two points are not used in subsequent analysis. The
lowest measurement of $\dot{\rho}_*$ at $z\approx 6$ is from \citet{Bun:04},
based on UDF dropouts, and probes to $0.1\,L^*$. Contributions from fainter
sources are unlikely to increase this measurement by more than a factor of two.
The measurement of $\dot{\rho}_*$ between these extremes comes from
\citet{Bou:05}, incorporating the largest current sample of $i$-dropouts
from the UDF, UDF parallels, and GOODS. This measurement supersedes an
earlier measurement \citep{Bou:04a} using only the UDF parallel fields,
which is not shown.

Although the issue of dust obscuration at high
redshift is still highly uncertain, some data is beginning to be obtained.
In addition to the $E(B-V)$ estimates from \citet{Ouch:04}, intriguing
evidence for significant obscuration ($A_V\approx 1\,$mag) at $z=6.56$ has
recently been established through Spitzer observations of a lensed Lyman
$\alpha$ (Ly$\alpha$) emitting source \citep{Cha:05}. This implies that the
first epoch of star formation in this source must have occurred around
$z\approx 20$, and moreover that the ``common" obscuration corrections applied
to UV luminosity based SFR densities at $3\lesssim z\lesssim 6$ may not be
unreasonable. This is also supported by spectroscopic Ly$\alpha$ emission
measurements of Lyman Break Galaxies (LBGs) at $z\approx 5$ \citep{And:05},
suggesting that the bright LBGs (at least) lie in dusty, chemically evolved
systems at this redshift.

At redshifts $6<z<10$ there have also been recent exciting estimates of
the SFH \citep{Bou:04b,Bou:05b,Bou:05c} based on UV luminosities
inferred using the photometric dropout technique. These measurements strongly
suggest that the decline in the SFH seen around $z\approx 6$ continues
to higher redshifts. Given the very small samples involved, the
complete uncertainty regarding the level of obscuration at these redshifts,
and more importantly the minimal impact that these data have on the
integrated properties of the SFH being explored here, we do not include these
points in any of our subsequent analysis. It is interesting to note, though,
that they do appear in general to be consistent with all our results based on
the SFH at $z\lesssim 6$.

To summarise the impact of dust obscuration corrections on the normalisation
of the SFH, first the corrections can obviously act only to increase,
not decrease, the normalisation. The technique of UV$+$FIR measurements gives
an effective obscuration correction to the UV data increasing from a factor of
two at $z\approx 0$ up to a factor of five at $z\gtrsim 1$. These results
are consistent with obscuration corrected H$\alpha$ estimates for
$\dot{\rho}_*$ spanning $0<z\lesssim 2$ (the range for which H$\alpha$
estimates are available), suggesting that the extent of the obscuration
correction is unlikely to be much smaller.
For $0.5 \lesssim z \lesssim 2.5$ this becomes less of a concern as the
SFH is dominated by contributions from FIR measurements, unaffected by
obscuration, which serve as a lower limit to the SFH normalisation.
At higher redshifts still, the issue is less clear, but the recent
results indicated above suggest that even up to $z\approx6$ dust obscuration
may be significant. For the following analysis, the $\overline{\nu}_e$ flux
is dominated by the $z\lesssim1$ regime, where the obscuration correction
seems fairly well constrained through the UV$+$FIR technique.

\subsection{IMF assumptions}
\label{imf}
While uncertainties in SFR calibration act to increase the scatter in
the SFH, and uncertainties in dust obscuration can raise it to greater
or lesser degrees, the choice of IMF is the only assumption that can
systematically {\em decrease\/} the SFH normalisation. While most authors
over the past decade have assumed the traditional \citet{Sal:55} IMF for
convenience, observations within recent years have strongly ruled it out as
a robust model for a universal IMF. A modified Salpeter form with a
turnover below $1\,M_{\odot}$, though, is still a reasonable model
\citep[e.g.,][]{Bal:03}, and other currently favoured IMFs include those of
\citet{Kro:01}, \citet{Bal:03}, and \citet{WK:05,WK:06}.

The SFR calibrations for all indicators used here for deriving the SFH
estimates, \citep[which are the same as in][]{Hop:04}, are defined assuming
the \citet{Sal:55} IMF. The compilation of data by \citet{Hop:04} converts
SFH estimates from the literature to these calibrations, where necessary,
to ensure consistent assumptions throughout. To convert SFH estimates
to an alternative IMF assumption corresponds to simple scale factor.
This scale factor is typically established through population synthesis
modelling, using codes such as PEGASE \citep{FR:97}, GALAXEV \citep{BC:03},
or STARBURST99 \citep{Lei:99}. These codes, given an input IMF and SFR,
can be used to infer the luminosity at SFR-sensitive wavelengths (typically
UV and H$\alpha$). For different IMFs, the ratio between the resulting
luminosities for a fixed SFR is the required scale factor. To explore
these scale factors we used PEGASE.2 to infer the UV (2000\,\AA)
and H$\alpha$ luminosities, given as input a fixed, constant SFR, for
those IMFs not excluded by \citet{Bal:03}. We use the default PEGASE.2
values for most input parameters, including a close binary mass
fraction of 0.05, evolutionary tracks with stellar winds, and the SNII
model B of \citet{WW:95}, but we specify an initial metallicity of
$Z = 0.02$, evolving self-consistently. We refer the reader to
\citet{BC:03} for a comparison between GALAXEV and PEGASE.2
\citep[and also][who explore the effect of different input
metallicities]{Far:06}. For IMFs with progressively shallower slopes,
it can be seen that the H$\alpha$ and UV luminosities scale differently,
compared to those derived assuming the \citet{Sal:55} IMF,
with H$\alpha$ being enhanced more quickly than the UV. We choose to
follow earlier work, and rely on the UV luminosity for our IMF scale factors
\citep[e.g.,][]{Col:01,Mad:98}, but note that our results would change
only marginally if we used the scale factors from the H$\alpha$ luminosities.

The scale factors derived for shallower (flatter, or more top-heavy) IMFs
are, in general, smaller than those for steeper IMFs. This results from the
top-heavy IMFs producing more high mass stars, and consequently more UV or
H$\alpha$ luminosity, for a fixed total mass or SFR. Conversely, for a fixed
UV or H$\alpha$ luminosity, a top-heavy IMF requires a lower SFR
to reproduce that luminosity. We emphasize that here the {\em observed}
quantity is the UV or H$\alpha$ luminosity, and the {\em derived} quantity is
the SFR. We investigate here two extreme, but still
realistic, IMF possibilities. The factor to convert SFH measurements from a
traditional \citet{Sal:55} IMF (with a power law slope of $-1.35$) to the IMF
of \citet[hereafter BG IMF, having a high mass power law
slope of $-1.15$]{Bal:03} is $0.50$ ($-0.305\,$dex). (Note that
\citeauthor{Bal:03} \citeyear{Bal:03} quote a high mass slope of $-1.2$
in their Table~2, from the local H$\alpha$ luminosity density,
but also quote $-1.15$ as the best-fit from the cosmic SFH. Here, as we
are interested in the extremes, we choose to use the latter.) If we used
the scaling from the PEGASE.2 H$\alpha$ luminosity, we would have a factor
of $0.41$ rather than $0.50$ for the BG IMF. All other IMFs explored here
vary much less in the relative scaling for the H$\alpha$ and UV luminosities.
To convert to the modified Salpeter A IMF
\citep[hereafter SalA IMF,][with high mass power law slope of
$-1.35$]{Bal:03} is a factor of 0.77 ($-0.114\,$dex). The \citet{Kro:01} IMF
(high mass power law slope of $-1.3$) and the modified Salpeter B IMF
\citep[hereafter SalB IMF,][high mass power law slope of $-1.35$]{Bal:03} have
scale factors intermediate between these choices. By exploring the impact of
assuming the two extreme IMF choices we expect to provide bounds encompassing
the result from choosing any reasonable IMF in our subsequent analysis.
We refer the reader to Figure~1 of \citet{Bal:03} for an illustration of
these and other various IMFs from the literature, and also to Table~1 of
\citet{Far:06}.

The application of an IMF in this manner neglects, of course, the
possibility that the IMF is not universal and indeed is even likely
to be evolving itself \citep[e.g.,][]{Kro:01}. Different assumed
star formation histories for our own galaxy may also affect
estimates for the local IMF \citep[see discussion by][]{ES:06}.
At any epoch, though, the universe has some average IMF, which may have a
large scatter around it for individual objects, and this average may vary with
epoch. It is this average IMF that the SFH is sensitive to, and this may be
different from what is inferred locally in the Milky Way. The current
measurements do not yet support any detailed exploration of these issues
and we ignore them for the current analysis.

\section{SFH fitting}
\label{fits}
In order to derive a $\overline{\nu}_e$ flux from the DSNB for comparison
with the limits from SK, it is necessary to fit some functional form to the
SFH in order to facilitate integration over redshift. We choose to use the
parametric form of \citet{Col:01} as is now commonly used by many authors:
$\dot{\rho}_*=(a+bz)h/(1+(z/c)^d)$, here with $h=0.7$. The individual
$\dot{\rho}_*$ measurements chosen to constrain this fit are also
important since the resulting fit will obviously vary depending on the
data used. For $z\le1$, the SFH now appears to be very tightly constrained by
the combination of UV data from SDSS \citep{Bal:05}, COMBO17 \citep{Wol:03},
and GALEX \citep{Arn:05}, corrected for obscuration using the Spitzer FIR
measurements of \citet{LeF:05}. This use of the FIR measurements of
\citet{LeF:05} is further supported by their high level of consistency with
those from \citet{Per:05} and the robust local 1.4\,GHz estimate from
\citet{Mau:05}. As a consequence we use only this set of corrected UV$+$FIR
measurements, along with the $z=0.01$ H$\alpha$ estimate of \citet{Han:06},
to constrain the parametric fit for $z<1$.
For $z>1$ we use all the data available in the compilation with exceptions
as noted above (the \citeauthor{Tho:06} \citeyear{Tho:06} UDF estimates,
the two highest estimates at $z\approx 6$, and the three high-$L$ only
estimates from \citeauthor{Bou:03a} \citeyear{Bou:03a}) and we further exclude
the six lowest measurements between $1<z<2$. The latter include the three
highest redshift [OII] estimates from \citet{Hog:98}, and three UV estimates,
where the ``common" obscuration correction assumed is likely to significantly
underestimate the true level of obscuration
\citep[compare the UV points between Figures~1 and 2 of][for example]{Hop:04}.
This is not unexpected as the UV luminosity density, in particular, at
this redshift probes a very small fraction of the total
$\dot{\rho}_*$ \citep[see also][]{Tak:05}.

The parametric fitting is a simple $\chi^2$ fit to the 58 selected
$\dot{\rho}_*$ measurements spanning $0 \le z \lesssim 6$.
The corresponding DSNB is calculated following the description in
\citet[their equations~1 and 2]{BS:06}. The $\dot{\rho}_*(z)$ is first
converted to a type~II supernova rate history, $\dot{\rho}_{\rm SNII}(z)$,
scaling by the appropriate integral over the IMF
\begin{equation}
\label{sniirate}
\dot{\rho}_{\rm SNII}(z) = \dot{\rho}_*(z)
 \frac{\int_8^{50}\psi(M)dM}{\int_{0.1}^{100} M \psi(M)dM}
\end{equation}
\citep[see, e.g.,][]{Dah:04,Mad:98}, where we have neglected the small delays
due to the short lifetimes of SNII progenitors. For the two IMFs explored here
we have $\dot{\rho}_{\rm SNII}(z)=(0.0132/M_{\odot})\,\dot{\rho}_*(z)$ for the
BG IMF and $\dot{\rho}_{\rm SNII}(z)=(0.00915/M_{\odot})\,\dot{\rho}_*(z)$
for the SalA IMF. This illustrates that the choice of IMF will affect the
derived neutrino production through two separate but related normalisations.
The first comes from how the IMF affects the normalisation of the derived
SFH from the observed UV, H$\alpha$, or other SFR-sensitive luminosity,
the second through this conversion of the SFH into a SNII rate history.
The combination of these two factors to some degree converge on similar
results, with steeper IMFs having higher SFH normalisations, but lower SNII
rate scalings, and vice-versa. As we show below the BG IMF produces SNII
rates that are a factor of $0.94 =(0.50/0.77)\times(0.0132/0.00915)$ of those
from the SalA IMF (see also additional discussion below in \S\,\ref{snii}).
We emphasise that, for a fixed UV, H$\alpha$, or other SFR-sensitive
luminosity, a shallower high-mass IMF produces fewer SNII, while a steeper
IMF produces more. If there are fewer SNII, this will respect the upper
limit on the DSNB flux, but may start to conflict with direct measurements of
$\dot{\rho}_{\rm SNII}$ (which we take to be a lower limit) if too few
SNII are predicted. The converse applies for the steeper IMFs, with more SNII
predicted.

It should be noted, too, that the SNII rate could be derived more directly
from (say) the UV luminosity, rather than through the calibration to an
SFR (representative of the whole mass range of an IMF), and then back to
a SNII rate in this way. We use this method as it conveniently allows
the confidence region fit to the SFH data to be used directly rather than
calculating different SNII rate conversions for each SFH measurement,
depending on the SFR-sensitive wavelength.
We have assumed throughout that all IMFs span the mass range
$0.1<M<100\,M_{\odot}$. Allowing a mass range up to $125\,M_{\odot}$ alters
most quantities by less than 1\% and all quantites by less than 2\%,
significantly less than the variation between different IMF choices, the
measurement uncertainties, or other uncertainties affecting the SFH
normalisation. The choice of stellar mass range that gives rise to SNII is
the largest assumption in this step. Restricting the upper mass limit
to $30\,M_{\odot}$ reduces the scale factor by about $10\%$ in both cases.
A much greater change is introduced by raising the lower mass limit. With
a mass range of only $10<M<30\,M_{\odot}$ in the numerator of
Equation~\ref{sniirate}, the resulting scale factors are reduced by
$\approx 40\%$.

The predicted differential neutrino flux (per unit energy) is then
calculated by integrating $\dot{\rho}_{\rm SNII}(z)$ multiplied by the
$\overline{\nu}_e$ emission per supernova, $dN/dE'$, appropriately redshifted,
over cosmic time \citep{FK:03,And:04,Str:04,Str:05,Dai:05,Lun:05}:
\begin{equation}
\frac{d\phi(E)}{dE} = c \int_{z=0}^6 \dot{\rho}_{\rm SNII}(z)
 \frac{dN(E(1+z))}{dE'} (1+z) \frac{dt}{dz} dz,
\end{equation}
where $dt/dz=(H_0(1+z)\sqrt{\Omega_M(1+z)^3+\Omega_\Lambda})^{-1}$,
$c$ is the speed of light, and we are evaluating the thermal emission
spectrum
\begin{equation}
\frac{dN}{dE'}(E')=\frac{E_\nu^{\rm tot}}{6} \frac{120}{7\pi^4} \frac{E'^2}{T^4}
\left[e^{E'/T}+1\right]^{-1},
\end{equation}
at the appropriately redshifted energy $E'=E(1+z)$
(note the choice of units where $k=1$ so that $T$ has units of energy).
Finally, $\int_{19.3\,{\rm MeV}}^{\infty} (d\phi/dE) dE$ is calculated
to establish the $\overline{\nu}_e$ flux for comparison with the SK limit
of 1.2\,cm$^{-2}$\,s$^{-1}$ \citep{Mal:03}.
The assumption of a thermal emission spectrum, and the associated
choices of temperature, are determined by the eventual decoupling
(after diffusion) of neutrinos from the hot and dense proto-neutron
star, at a radius called the ``neutrinosphere"~\citep{Raf:96}.
We explore the implication of assuming a temperature of
$T\approx4\,$MeV, 6\,MeV, or 8\,MeV, and as in \citet{BS:06}, we assume
$E_\nu^{\rm tot}=3\times10^{46}\,$J$=3\times10^{53}\,$erg for the total energy
carried by all six neutrino flavours. While supernovae emit all flavours
of neutrinos and antineutrinos, and each is assumed to carry an approximately
equal fraction of the total energy, at present the $\overline{\nu}_e$ flavour
is the most detectable. Because the temperatures of the muon
($\overline{\nu}_{\mu}$) and tauon ($\overline{\nu}_{\tau}$) antineutrinos are
expected to be higher than for $\overline{\nu}_e$, the effect of neutrino
mixing may be to increase the observable $\overline{\nu}_e$ flux for a given
SFH \citep{FK:03,And:04,Str:04,Dai:05}; as in \citet{BS:06} and \citet{Yuk:05}
our results apply to an effective temperature after mixing, making them
more constraining.

Given the $\overline{\nu}_e$ flux for each temperature, we simply
scale the best fitting SFH to ensure the SK limit is not violated.
This approach has the advantage that the shape of the SFH is determined only
by measured data that is differential in redshift, while the normalisation
comes directly by imposing the SK $\overline{\nu}_e$ limit.
We also explored an alternative approach that did not restrict the fitted
shape of the SFH. In this approach, while iterating over the four fitting
parameters, each new potential minimum $\chi^2$ SFH is subjected to the
DSNB limit, and the fit is rejected if the limit is violated. This technique
resulted, for the higher $\overline{\nu}_e$ temperatures, in fitted
SFHs that were significantly low (compared to the measurements) in the
mid-to-high redshift range, while still matching the $z=0$ and $z\gtrsim 4$
SFH. For the low $\overline{\nu}_e$ temperatures, this method gave identical
results to the original approach. Our preferred approach emphasises the
direct impact of the $\overline{\nu}_e$ data appropriately on
the $z \lesssim 1$ measurements, as there are very few detectable neutrinos
received from higher redshift. The best fit SFHs independent of the
$\overline{\nu}_e$ limit are identical to the fits constrained by a
$\overline{\nu}_e$ temperature of $T=4\,$MeV. As found and discussed in
\citet{Yuk:05}, our results favour effective temperatures at the lower end of
the predicted range \citep{Kei:03,Tho:03,Lie:05,Sum:05}.

In addition to the \citet{Col:01} parameterisation, we also explored a
piecewise linear SFH model in $\log(1+z)-\log(\dot{\rho}_*)$ space,
in order to test the possibility that the \citet{Col:01} parametric model
could be biasing the shape of the resulting SFH fit in some way. In this model
we allow the following six parameters to vary: The $z=0$ intercept, the slopes
of three linear segments and the two redshift values at which the slopes
change. With this model we similarly explore the range of $\overline{\nu}_e$
temperatures, as for the \citet{Col:01} parameterisation above.

\section{Results}
\label{results}
\begin{figure*}
\centerline{\rotatebox{-90}{\includegraphics[width=7.0cm]{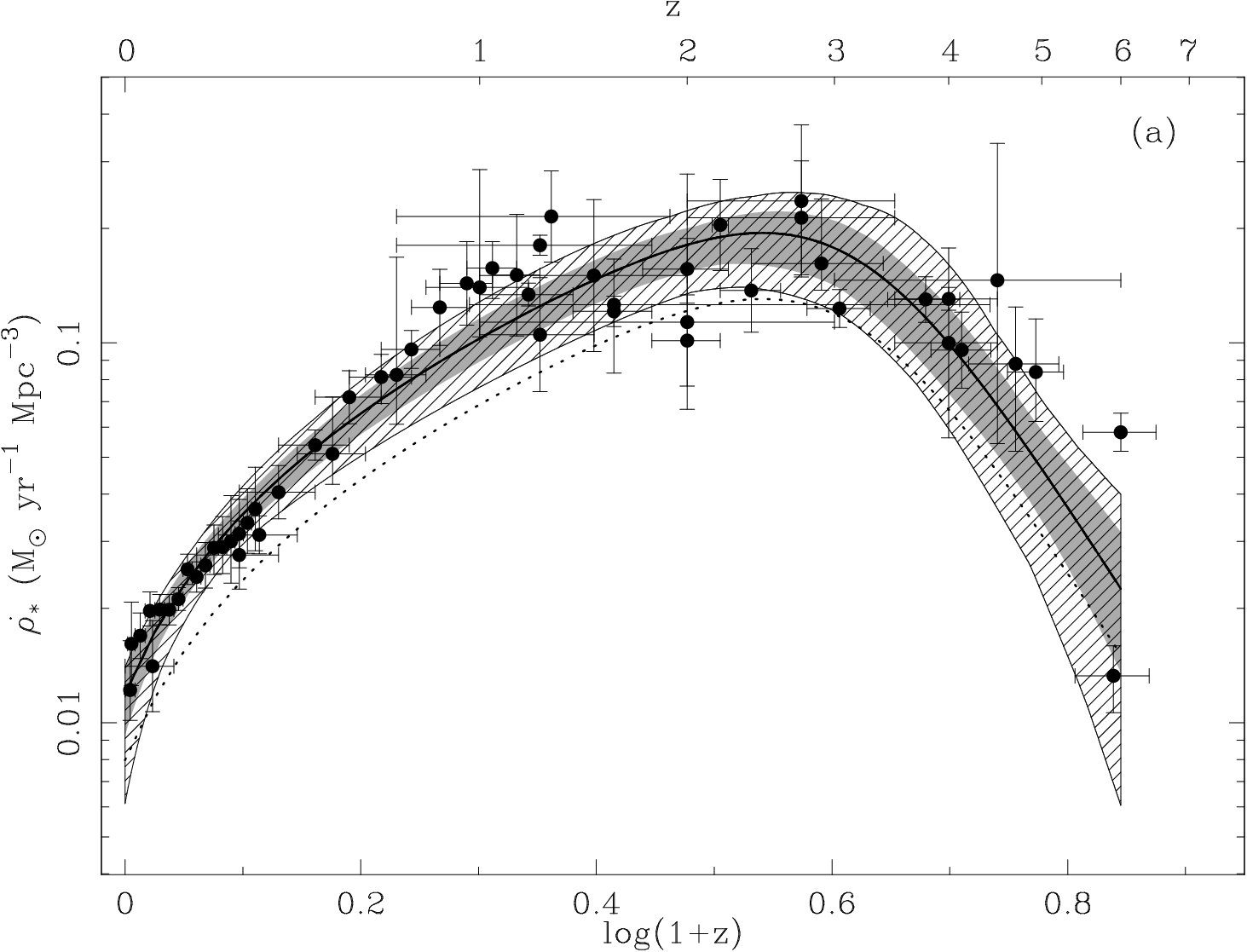}}
\rotatebox{-90}{\includegraphics[width=7.0cm]{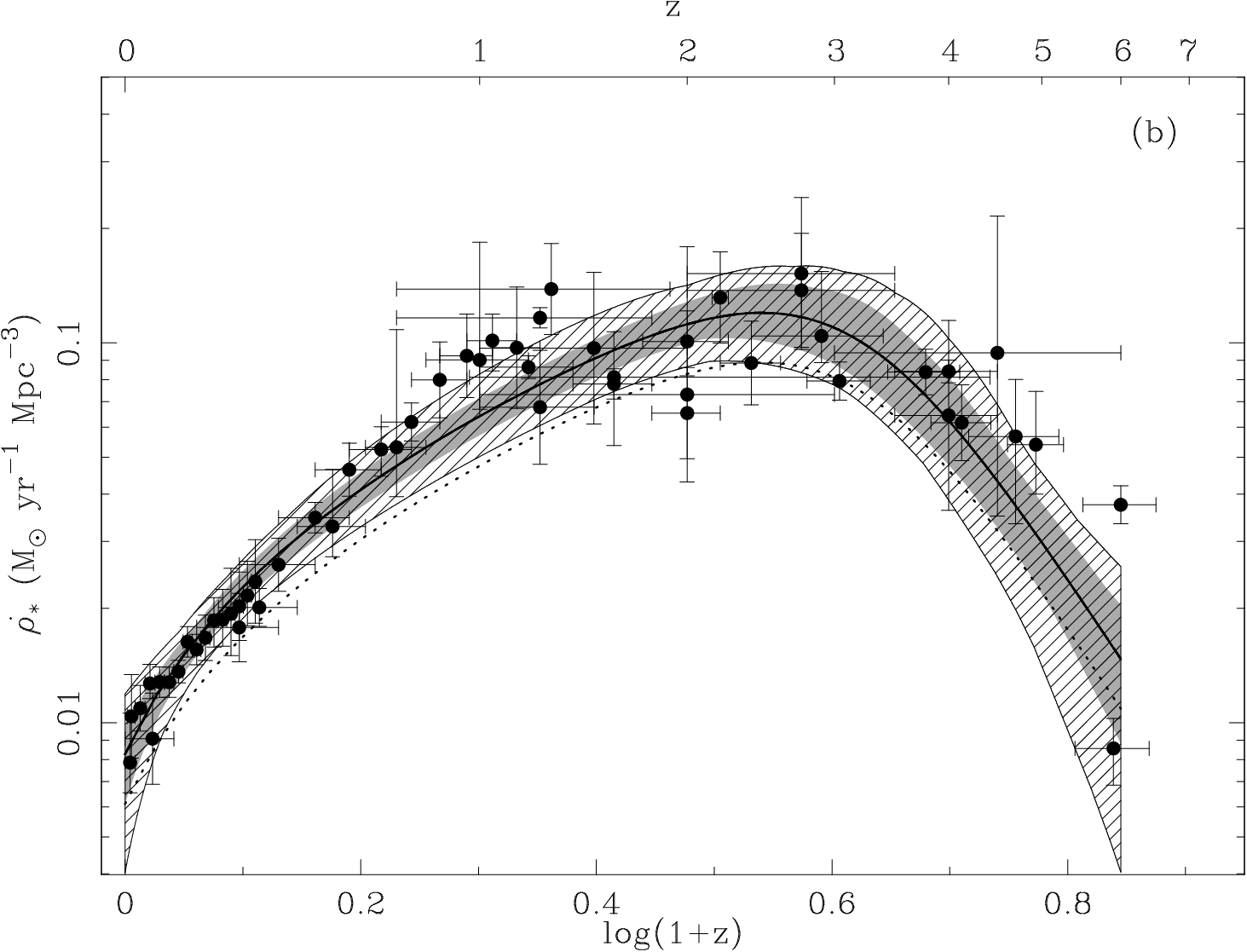}}}
\caption{SFR density data used in defining the best fitting parametric
forms, and the resulting fits. (a) Assumes SalA IMF. (b) Assumes BG IMF.
The shape of the fits is determined from the SFH data alone, and a scaling
factor is fit to ensure consistency with the SK $\overline{\nu}_e$ limit
(i.e., given the assumed temperature, this quantifies how much lower the
SFH normalisation has to be so as not to violate the SK limit).
Solid lines assume a $\overline{\nu}_e$ temperature of $T=4\,$MeV
or $T=6\,$MeV, and dotted $T=8\,$MeV. The grey shaded and
hatched regions are the $1\,\sigma$ and $3\,\sigma$ confidence regions around
the $T=4\,$MeV fits respectively. The scaling factors are:
(a)~$(1.0,1.0,0.67)$ and (b)~$(1.0,1.0,0.74)$ respectively for
$T=(4,6,8)\,$MeV.
 \label{fig:sfhfit}}
\end{figure*}

Figure~\ref{fig:sfh} shows the current SFH data compilation (assuming the
SalA IMF) emphasising the additional data used in this analysis compared to the
compilation of \citet{Hop:04}\footnote{The $\dot{\rho}_*$ data from
\citet{Hog:98} as given by \citet{Hop:04} are incorrect, a result of an error
in the cosmology conversion parameters used in that analysis. The correctly
converted data are shown here, and are smaller than those given by
\citet{Hop:04} by values decreasing monotonically from $\approx 43\%$ for the
$z=0.2$ bin to $\approx 30\%$ for the $z=1.2$ bin. Sincere thanks go to
Chun Ly for bringing this error to our attention.}.
The best fitting \citet{Col:01} form for this IMF is also shown
($a=0.0170$, $b=0.13$, $c=3.3$, $d=5.3$), as is the best-fitting piecewise
linear fit. Figure~\ref{fig:sfhfit} shows the data used in the fitting and
the best fits assuming three temperature values for the $\overline{\nu}_e$
population for each IMF assumed. The \citet{Col:01} parameters for
each case are given in Table~\ref{tab1}. With 58 data points and 4 free
parameters there are 54 degrees of freedom, and the best-fitting $\chi^2$
values (for the $T=4\,$MeV cases) are 37.5 for both the SalA IMF and
the BG IMF. This value is perhaps somewhat lower than expected, and reflects
the nature of the uncertainties for the heterogeneous data used in the fit.
These come primarily
from the Poisson counting statistics of the numbers of observed galaxies
contributing to each measurement. No attempt has been made to resample
or reanalyse independent data to appropriately combine their uncertainties,
and the effect of multiple independent measurements at similar redshift,
each with independently calculated uncertainties (since the data are
heterogeneous), is to mimic conservative uncertainty estimates in a
homogeneous dataset. Correlations may also exist between assumed
independent SFH measurements at similar redshift, which may also
contribute to the low $\chi^2$ value. This may arise as a result, for
example, of different teams independently analysing the same dataset,
or of independent analyses of common or overlapping survey areas, either
at a range of wavelengths within the same survey, or from surveys at
different wavelengths of the same region of sky.

\begin{figure*}
\centerline{\rotatebox{-90}{\includegraphics[width=9.0cm]{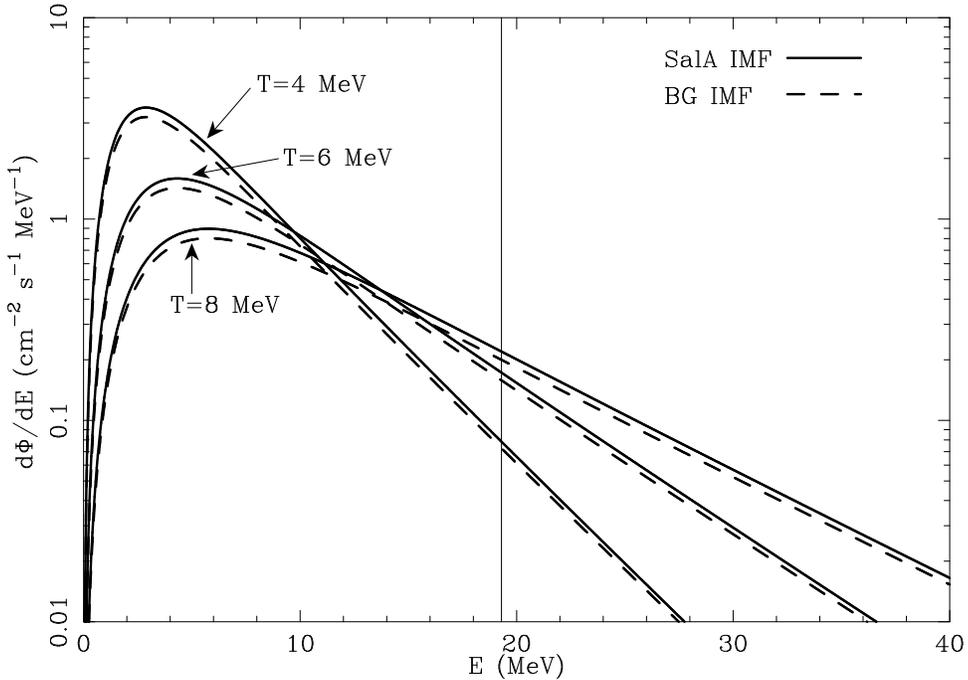}}}
\caption{The $\overline{\nu}_e$ energy spectra predicted for the
various SFH fits and temperature assumptions. The solid and dashed
curves are the SalA IMF and BG IMF respectively. The $T=4\,$MeV and
$T=6\,$MeV curves are consistent with the SK $\overline{\nu}_e$ limit.
The $T=8\,$MeV curves are inconsistent with the $\overline{\nu}_e$ limit,
and indicate the shape of the $\overline{\nu}_e$ spectrum derived by assuming
the parametric form for the SFH corresponding to our best fit ($T=4\,$MeV),
and setting the $\overline{\nu}_e$ temperature to the higher value.
The thin vertical line marks $E=19.3\,$MeV, above which the
$\overline{\nu}_e$ contribute to the SK limit.
 \label{fig:sneas}}
\end{figure*}

For both assumed IMFs it can clearly be seen that the assumption of
$T=8\,$MeV, when the SFH is required to be consistent with the
$\overline{\nu}_e$ flux limit, is inconsistent with the SFH measurements.
Also for both IMFs, the best fitting SFH assuming $T=6\,$MeV is identical to
that assuming $T=4\,$MeV. This can be understood by considering the SalA IMF,
for example, with the higher SFH normalisation, but which also has a lower
conversion factor between $\dot{\rho}_*$ and $\dot{\rho}_{\rm SNII}$, causing
the predicted $\overline{\nu}_e$ flux to be within the SK limit even with
the assumption of the slightly higher neutrino temperature. For both IMF
assumptions we determine the $1\,\sigma$ (grey-shaded) and
$3\,\sigma$ (hatched) confidence regions around the best fitting SFH
(corresponding to $T=4$ or $6\,$ MeV). These are derived from
the regions of parameter space with $\chi^2<\chi^2_{\rm min}+\Delta\chi^2$
where $\Delta\chi^2=4.7$ and $16.0$ respectively for $1$ and $3\,\sigma$
\citep[see][who shows that $\Delta\chi^2$ for $q$ ``interesting" parameters
itself follows a $\chi^2$ distribution with $q$ degrees of freedom; here
$q=4$ for the \citeauthor{Col:01}\ \citeyear{Col:01} parameterisation]{Av:76}.
These confidence regions are determined independently at each redshift,
intended to encompass the envelope of all fits. They are thus likely to
be somewhat conservative, an effect that is compounded by the result of the
$\chi^2$ analysis being affected by what are effectively conservative
measurement uncertainties. Subsequent Figures reproduce
these confidence regions in the predictions for stellar and metal mass density
evolution ($\rho_*(z)$ and $\rho_Z(z)$, respectively) and SN rate evolution
($\dot{\rho}_{\rm SN}(z)$).
The shape of the $\overline{\nu}_e$ energy spectra corresponding to the
various fits for the different IMF and temperature assumptions are shown
in Figure~\ref{fig:sneas}. These results are very similar to the spectral
shapes derived in Figure~1 of \citet{BS:06}.

\begin{figure*}
\centerline{\rotatebox{-90}{\includegraphics[width=7.0cm]{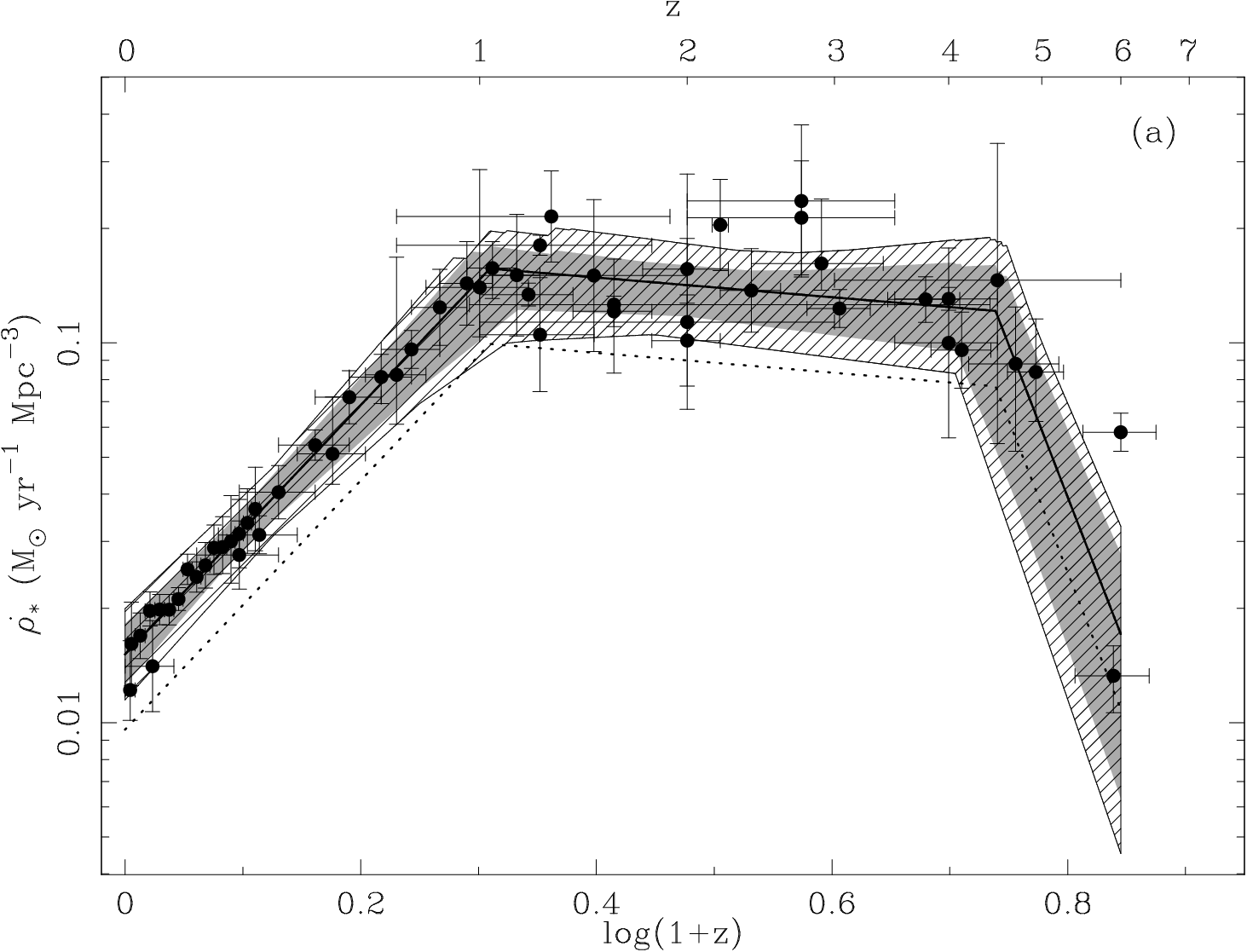}}
\rotatebox{-90}{\includegraphics[width=7.0cm]{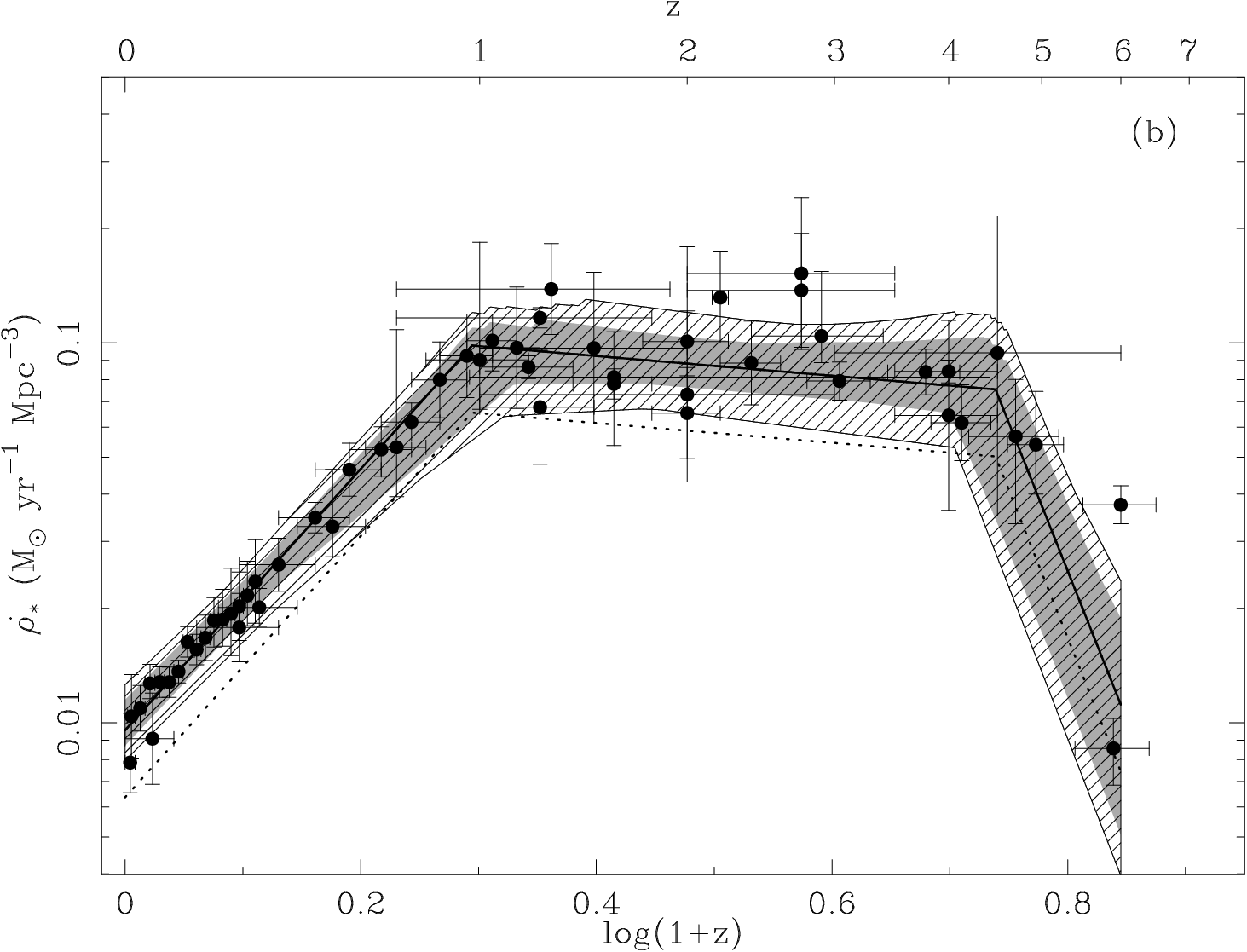}}}
\caption{As for Figure~\ref{fig:sfhfit} but assuming a piecewise linear
SFH model. (a) Assumes SalA IMF. (b) Assumes BG IMF. As before,
solid lines assume a $\overline{\nu}_e$ temperature of $T=4\,$MeV or
$T=6\,$MeV, and dotted $T=8\,$MeV.
 \label{fig:lines}}
\end{figure*}

The results of the piecewise linear fitting, seen in Figure~\ref{fig:lines}
and detailed in Table~\ref{tab:lines}, are remarkably similar in
general properties to the results in Figure~\ref{fig:sfhfit}. The grey
shaded and hatched regions here again show, respectively, the
$1\,\sigma$ and $3\,\sigma$ confidence regions, here corresponding to
$\Delta\chi^2=7.0$ and $19.8$ (with $q=6$). In this case the perimeters
of the confidence regions are not smooth, an artifact arising from a
combination of both the finite sampling of parameter space for which
$\chi^2$ values are calculated, and the linear nature of the parameterisation.
With sufficiently high resolution sampling of parameter space the
confidence region perimeters would be expected to curve more smoothly
around the upper left and right corners. This artifact does not impact
upon any of our analysis, and as the confidence regions are mainly shown
to illustrate the range of uncertainty in the fits, we do not pursue this
issue further. The piecewise linear fits again show low values
of $\chi^2$ (see Table~\ref{tab:lines}) for the same reasons as given
above. The same preference for lower $\overline{\nu}_e$ temperatures
is seen. The similarities here with the
\citet{Col:01} parameterisation are encouraging and suggest that the
the \citet{Col:01} parameterisation has not introduced any significant
bias against specific SFH shapes.
For subsequent analysis we retain the $T=4\,$MeV fits using the
\citet{Col:01} parameterisation. This does not affect our conclusions,
which remain unchanged regardless of which SFH parameterisation we choose.

\begin{figure*}
\centerline{\rotatebox{-90}{\includegraphics[width=6.5cm]{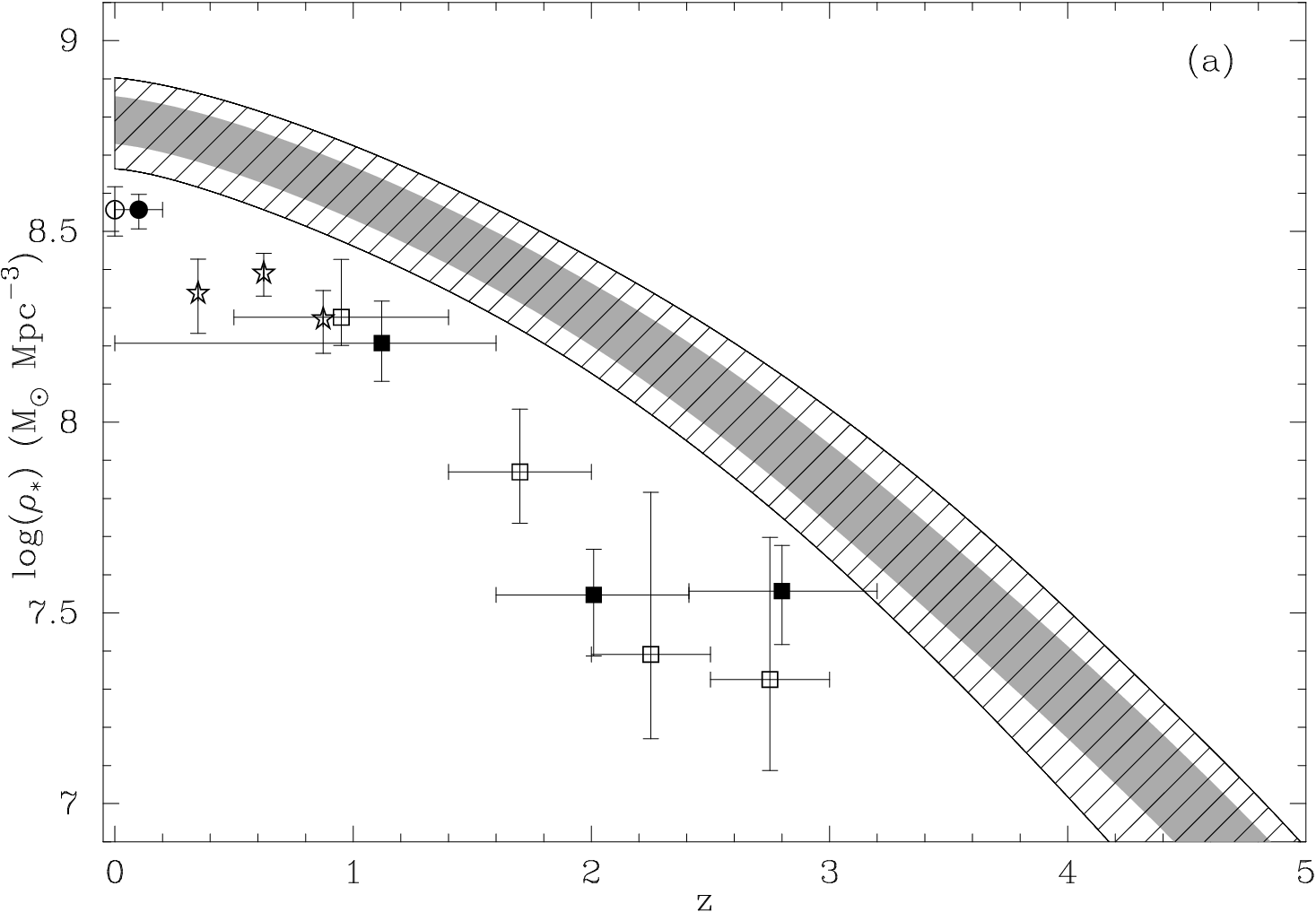}}
\rotatebox{-90}{\includegraphics[width=6.5cm]{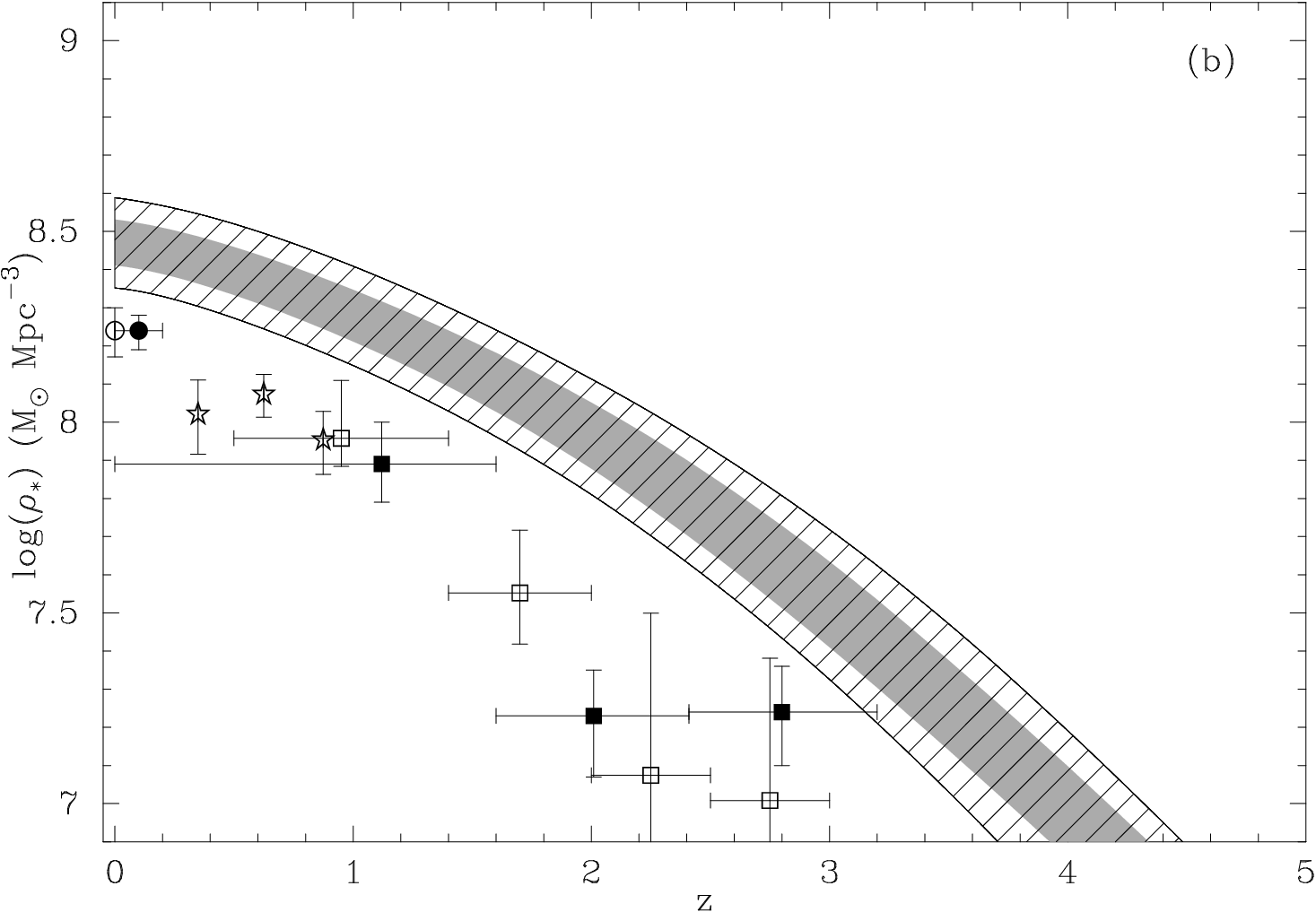}}}
\caption{Evolution of stellar mass density buildup inferred from the SFH.
(a) Assumes SalA IMF (with $R=0.40$). (b) Assumes BG IMF (with $R=0.56$). The
grey shaded and hatched regions come from the $1\,\sigma$ and $3\,\sigma$
confidence regions around the SFH $T=4\,$MeV fits respectively. The details
of scaling the data points to our assumed IMFs are given in the text. The open
circle is the local stellar density from \citet{Col:01}; the filled circle
and filled squares represent the SDSS and FIRES data, respectively, from
\citet{Rud:03}, scaled such that the SDSS measurement is consistent with
that from \citet{Col:01}; the open stars are from \citet{Bri:00}; and
the open squares are from \citet{Dic:03}.
 \label{fig:stars}}
\end{figure*}

Figure~\ref{fig:stars} shows the evolution of the stellar mass density,
$\rho_*(z)$, along with the predictions from the best fitting SFH for the two
extreme IMF assumptions \cite[compare with the extensive compilation of data
in Figure~4 of][]{Far:06}. To construct this diagram we need to know the
fraction of the stellar mass recycled into the interstellar medium as stellar
winds or SN ejecta, $R$, corresponding to each IMF
\citep{Col:01,Mad:98,Ken:94}. We follow the prescription suggested by
\citet{Col:01}, using the models of \citet{RV:81} and \citet{WW:95} for mass
loss due to stellar winds and supernovae respectively, and calculate
$R=0.40$ for the SalA IMF, and $R=0.56$ for the BG IMF. The stellar mass
inferred is then a fraction $(1-R)$ of the time integral of the SFH
\citep{Col:01}. Converting the observed stellar mass density measurements
(where a Salpeter IMF is most commonly used) to our assumed IMFs is achieved
by scaling by the product of the SFR conversion factor and the ratio of the
$1-R$ factor for the chosen IMF to that of the Salpeter IMF (where $1-R=0.72$).
The reliability of this method was confirmed by reproducing the stellar mass
estimate assuming the \citet{Ken:83} IMF by \citet{Col:01} compared to
their Salpeter IMF value. As a point of reference, using
the \citet{Sal:55} IMF where these scalings are not required, a diagram very
similar to the current Figure is shown in Figure~3 of \citet{Hop:05}.

Both plots in Figure~\ref{fig:stars} show the measurements lying
systematically below the predictions from the SFH, although the difference
becomes more significant at higher redshift ($z\gtrsim 1.5$). Causes for the
apparent inconsistency at high redshift have been discussed by other authors
\citep[e.g.,][]{Nag:04,Hop:05}, who suggest that, in this regime,
the observations might be missing up to half the stellar mass density.
We discuss this, and the low redshift discrepancy, further in \S\,\ref{disc}.

\begin{figure*}
\centerline{\rotatebox{-90}{\includegraphics[width=6.5cm]{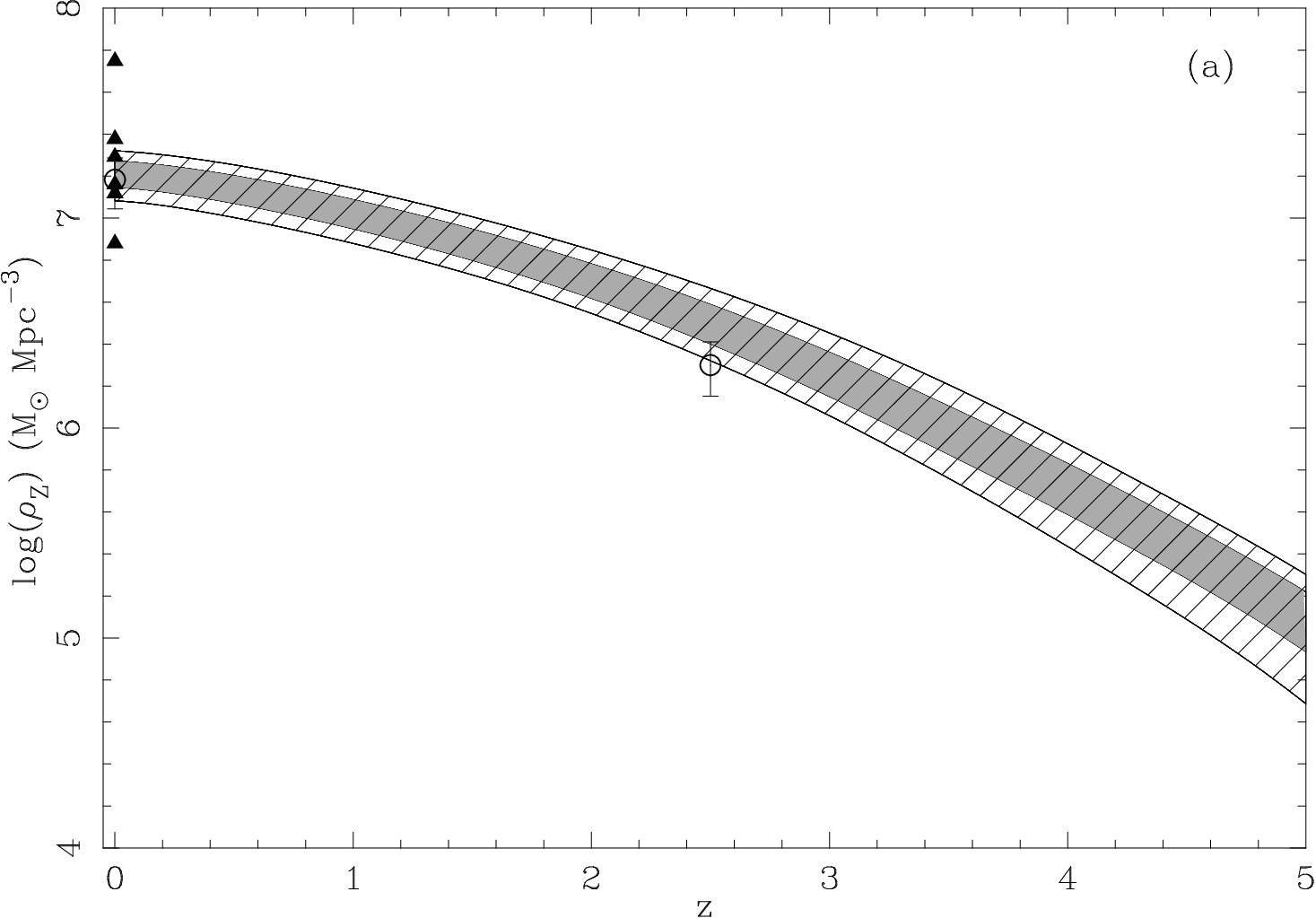}}
\rotatebox{-90}{\includegraphics[width=6.5cm]{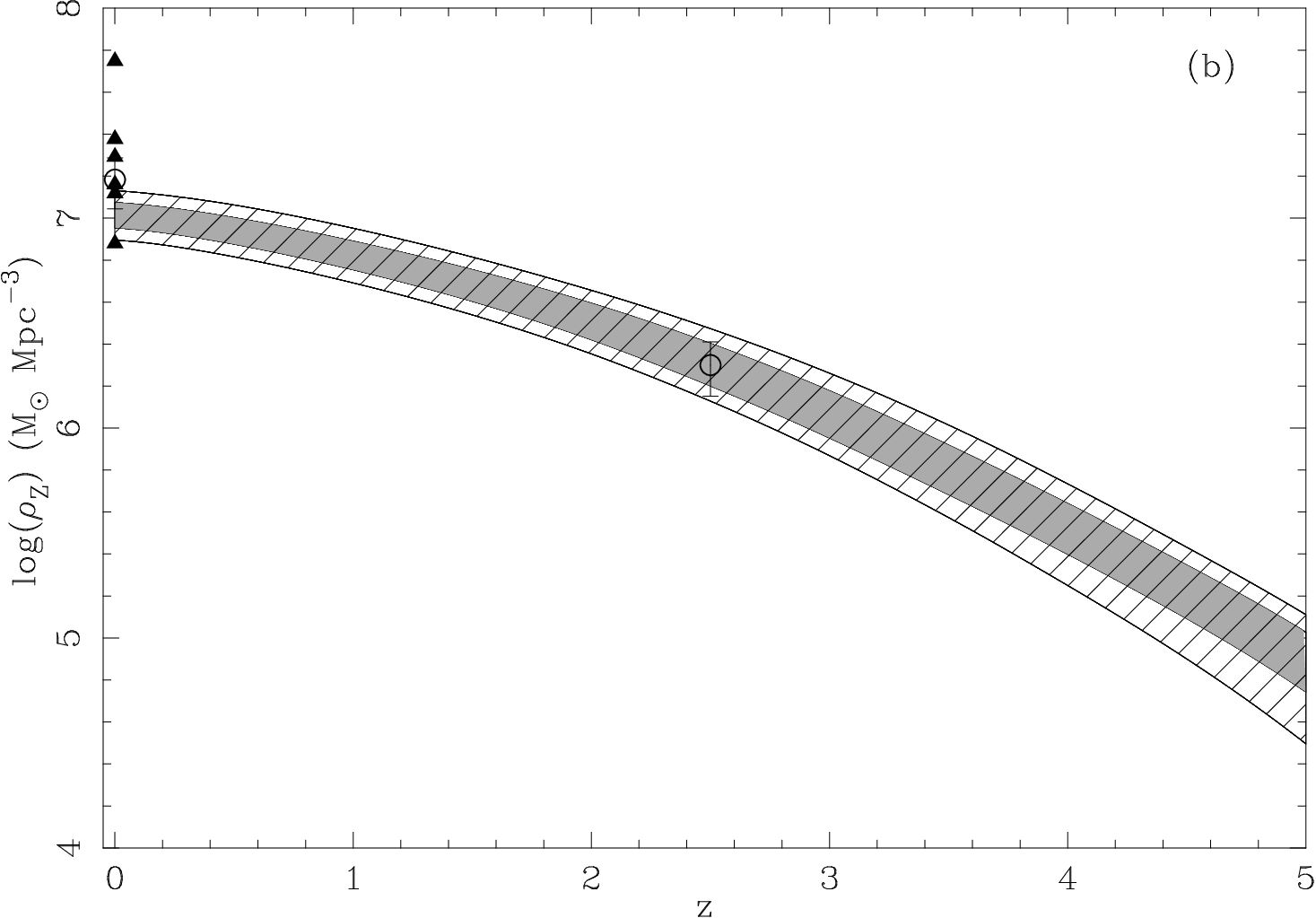}}}
\caption{Evolution of metal mass density buildup inferred from the SFH.
(a) Assumes SalA IMF. (b) Assumes BG IMF. The grey shaded and hatched
regions come from the $1\,\sigma$ and $3\,\sigma$ confidence regions around
the SFH $T=4\,$MeV fits respectively. The triangles at $z=0$ are from
\citet{CM:04}, and the open circles at $z=0$ and $z\approx 2.5$ are
from \citet{Dun:03}.
 \label{fig:metals}}
\end{figure*}

Figure~\ref{fig:metals} shows how the metal mass density evolves, $\rho_Z(z)$,
as inferred from the SFH \citep{PF:95,Mad:96}.
(Compare with \citeauthor{Hop:05}\ \citeyear{Hop:05}, and
for a more detailed treatment of the evolution of
separate metals, see \citeauthor{Dai:04}\ \citeyear{Dai:04}.)
To determine this relation from the SFH, we assume that
$\dot{\rho}_*=63.7\,\dot{\rho}_Z$ \citep[e.g.,][]{Con:03}. At $z=0$ the
compilation of data from \citet{CM:04} is shown, and these authors favour a
value of
$1.31\times10^7\,M_{\odot}$Mpc$^{-3}$, toward the low end of the
range. Values at $z=0$ and $z=2.5$ from \citet{Dun:03} are also shown,
suggesting that the evolution in $\rho_Z$ from the SFH may be
consistent with that estimated from the dusty submillimeter galaxy (SMG)
population, although recent results from \citet{BLP:05} indicate that the
SMGs may contribute much less to the metal mass density at high redshift.

\begin{figure*}
\centerline{\rotatebox{-90}{\includegraphics[width=7cm]{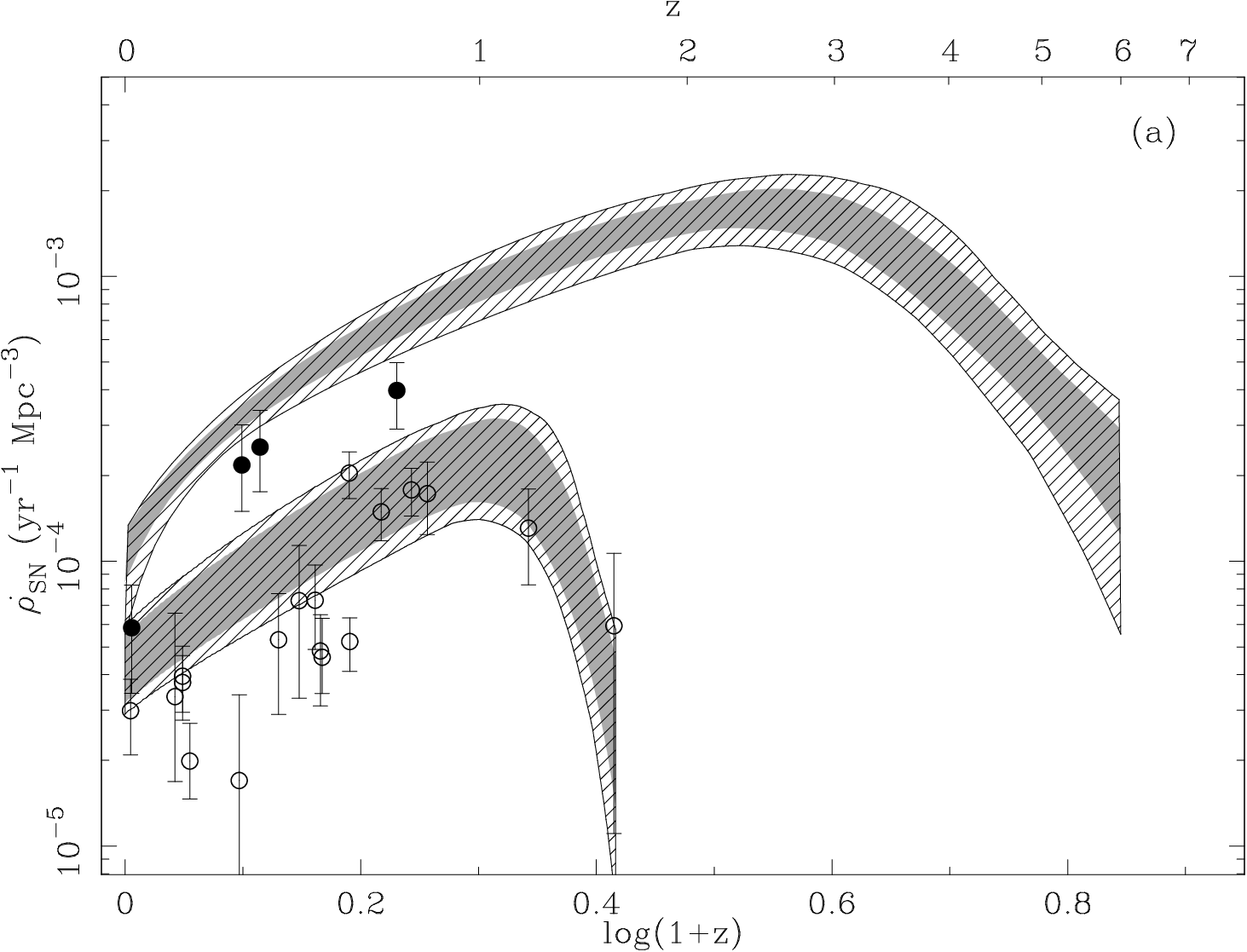}}
\rotatebox{-90}{\includegraphics[width=7cm]{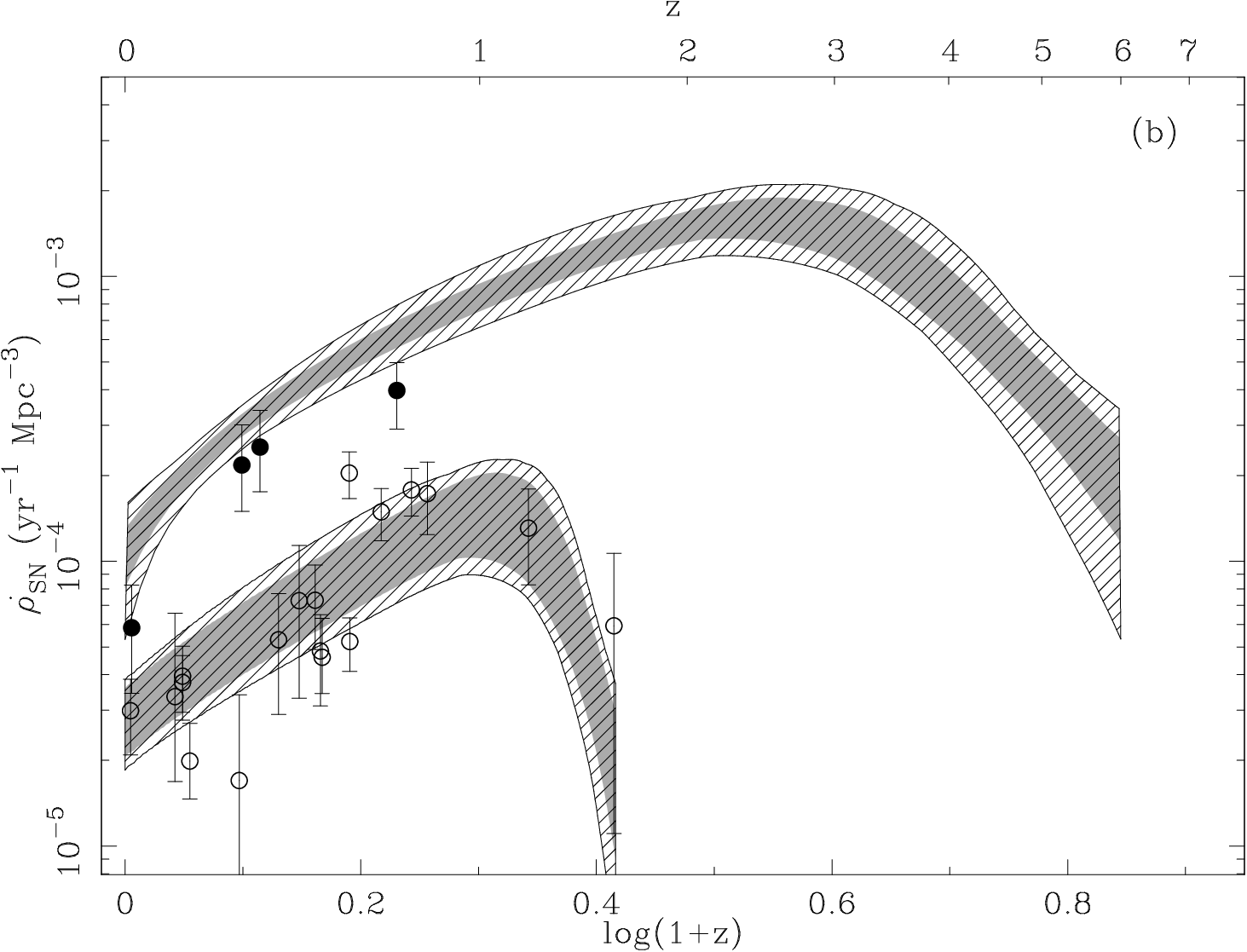}}}
\caption{Evolution of SN rates inferred from the SFH. The upper curves
correspond to the predictions for $\dot{\rho}_{\rm SNII}$, and the lower for
$\dot{\rho}_{\rm SNIa}$, assuming a delay time $t_{Ia}=3\,$Gyr.
(a) Assumes SalA IMF. (b) Assumes BG IMF. The grey
shaded and hatched regions again come from the $1\,\sigma$ and $3\,\sigma$
confidence regions around the SFH $T=4\,$MeV fits respectively, except that
for the SNIa region, the lower bound comes from assuming
$f_{Ia}=1/1000\,M_{\odot}^{-1}$, while the upper bound assumes
$f_{Ia}=1/700\,M_{\odot}^{-1}$. The filled circles are
$\dot{\rho}_{\rm SNII}$ measurements from \citet{Dah:04} and \citet{Cap:05}.
The open circles are $\dot{\rho}_{\rm SNIa}$ measurements reproduced from the
compilation of \citet{SR:06} and the data of \citet{Bar:06} and
\citet{Nei:06}.
 \label{fig:snrate1}}
\end{figure*}

Figure~\ref{fig:snrate1} shows the evolution in the SN rate for both types Ia
and II SNe. Although the data to date are not yet precise, the SN rate data
have a significant advantage over the stellar and metal mass density in that
they are differential in redshift, and are in principle more straightforward
to measure. The SNII rate density, $\dot{\rho}_{\rm SNII}$, is calculated from
the SFH as described above in Equation~\ref{sniirate}. The SNIa rate density,
$\dot{\rho}_{\rm SNIa}$, is similarly estimated, although it involves more
assumptions about the properties of SNIa events than in the case of SNII.
In particular the delay time $t_{Ia}$ between star formation and the SNIa
event and the efficiency $\eta$ of producing an SNIa event from objects in
the stellar mass range $3<M<8\,M_{\odot}$ are not well constrained (and even
this mass range is somewhat uncertain). Current estimates put $t_{Ia}$ roughly
in the range of $1-3$\,Gyr and $\eta$ of order $1-5\%$ \citep[see discussion
in][and references therein]{Str:05}. We follow \citet{Str:05} in
assuming a constant $t_{Ia}=3\,$Gyr, and $f_{Ia}=1/700\,M_{\odot}^{-1}$,
where $f_{Ia}=\eta \int_3^8 \psi(M)dM / \int_{0.1}^{100} M \psi(M)dM$,
to determine $\dot{\rho}_{\rm SNIa}$ from
\begin{equation}
\dot{\rho}_{\rm SNIa}(t)=\eta
\frac{\int_3^8\psi(M)dM}{\int_{0.1}^{100}M\psi(M)dM} \dot{\rho}_*(t-t_{Ia}).
\end{equation}
For the SalA and BG IMFs, $f_{Ia}= 0.028 \eta$ and $f_{Ia} = 0.032 \eta$,
respectively. With our assumed value of $f_{Ia}$ above, this corresponds to
assuming $\eta \approx 5\%$ for both IMFs.
Figure~\ref{fig:snrate1} also shows the effect of assuming
$f_{Ia}=1/1000\,M_{\odot}^{-1}$ in the lower limit of the $1\,\sigma$
and $3\,\sigma$ envelopes for the predicted $\dot{\rho}_{\rm SNIa}$.
Figure~\ref{fig:snrate2}a reproduces Figure~\ref{fig:snrate1}b with the
assumption of $t_{Ia}=1\,$Gyr, illustrating the effect of the different delay
times. This produces a somewhat reduced $\dot{\rho}_{\rm SNIa}$ envelope at
lower redshifts and moves the turnover to higher redshifts.

\section{Discussion}
\label{disc}

\subsection{Stellar mass and metal mass densities}

The predictions from the SFH for both $\rho_*(z)$ and $\rho_{Z}(z)$ are
difficult to analyse, for different reasons. Most predictions of
$\rho_*(z)$ based on SFH measurements seem to be larger than the observed
stellar mass density at high redshift, and numerous simulations imply that
the measurements might be underestimating the total $\rho_*(z)$
\citep[e.g.,][]{Nag:04,Men:04,Som:01,Gra:00}. Indeed it is suggested by
\citet{Dic:03} that additional obscuration added to their maximally old
component used in estimating stellar masses could cause arbitrarily large
masses to be derived, and it is perhaps not unreasonable to expect about a
factor of two larger stellar mass densities as a result of reasonable
obscuration levels. This would bring the measurements more into line with the
predictions from the SFH. Similar issues may affect the other high redshift
measurements of $\rho_*(z)$, and other issues that have also been raised
include incomplete galaxy population sampling and cosmic variance that may
affect surveys probing small fields of view
\citep[see discussion in][]{Nag:04}.

At low redshift, the discrepancy between the measurements of $\rho_*(z)$ and
the SFH prediction is more of a concern. A first attempt at resolving this
might be to suggest that the measured SFH is too high at $z=0$, and the
technique used (combining $\dot{\rho}_*$ from FIR and UV estimates) is not
accurate. This cannot be the whole solution as it does not address the problem
at $z\approx 1$, where the SFH is dominated by the FIR contribution,
and the $\rho_*(z)$ values inferred from the SFH are similarly higher than
the measurements. Another point against this solution is the equivalent
diagram in Figure~3 of \citet{Hop:05}, where (for a Salpeter IMF) a broader
region, encompassing the majority of SFH measurements, is used to predict
$\rho_*(z)$ rather than the confidence regions fitted here. Even in that
diagram, the $\rho_*(z)$ measurement at $z=0$ lies at the extreme lower
boundary of the SFH prediction. This suggests that there is something more
subtle underlying this discrepancy than simple obscuration correction errors
in the SFH. A partial solution might be found through the underlying
measurement techniques used for the different quantities. The SFH measurements
rely on inferring SFRs from the luminosity generated by massive stars, while
the $\rho_*(z)$ measurements come by inferring total stellar masses based on
the luminosity of low-mass stars. These are connected in the prediction
of $\rho_*(z)$ from the SFH through the assumed IMF shape, and it is possible
that the discrepancy seen in Figure~\ref{fig:stars} may be reflecting
limitations in our understanding of the relative shapes of the low and high
mass ends of our assumed IMFs. Although further exploration of this discrepancy
is beyond the scope of the present investigation, we refer the reader to
\citet{Far:06} who provide a detailed analysis of this issue, incorporating
limits from the total extragalactic background radiation along with
the SFH and stellar mass evolution.

Regarding the evolution of the metal mass density, $\rho_{Z}(z)$,
investigation of predictions from the SFH are complicated by the limited number
of estimates for this quantity at $z>0$. This is observationally a difficult
measurement to make, particularly as much of the metals may exist in an
ionised intergalactic medium component. Measurements of the contribution
from various components (stellar, dust, gas) have been explored with varying
estimates for how much of the metal mass density budget might be contained
in different components at high redshift \citep[e.g.,][]{Dun:03,BLP:05}.
Additional discussion of this issue, emphasising the (minimal)
contribution from damped Ly$\alpha$ absorbers, is presented by \citet{Hop:05}.

\subsection{Supernovae type~II}
\label{snii}

Almost identical predictions for $\dot{\rho}_{\rm SNII}$ are obtained for both
the SalA and SalB IMFs detailed in \citet{Bal:03}, a consequence of SalA
having a slightly higher conversion factor than SalB for transforming
$\dot{\rho}_*$ from the traditional Salpeter IMF, and a slightly lower
conversion factor to transform between $\dot{\rho}_*$ and
$\dot{\rho}_{\rm SNII}$. The BG IMF result (Figure~\ref{fig:snrate1}b) appears
to be marginally more consistent with both the $\dot{\rho}_{\rm SNII}$ and
$\dot{\rho}_{\rm SNIa}$ measurements than the prediction assuming the SalA IMF
(Figure~\ref{fig:snrate1}a) although both are acceptable. The uncertainties
affecting the $\dot{\rho}_{\rm SNII}$ measurements are treated in some detail
by \citet{Dah:04}, and the error bars shown are representative of the
combination of both statistical and systematic uncertainties. The $z=0$
measurement may be somewhat low, and recent investigations
\citep{Man:03, And:05a} suggest that this point is in fact likely to be higher
by a factor of two or three, consistent with the SFH predictions.
\citet{Dah:04} indicate that the main concern in the measurements is the level
of obscuration, and that a change from their assumed $E(B-V)=0.15$ of
$\Delta E(B-V)=\pm0.06$ would alter their measurements by one standard
deviation. It is certainly likely that this is the case for the highest
redshift point at $z=0.7$, where obscuration is expected to be higher than at
lower redshifts \citep[e.g.,][see also Figure~\ref{fig:sfh}]{Tak:05,Hop:04}.
This would have the effect of raising the $z=0.7$ measurement to be consistent
with the predictions from the SFH.

The $\dot{\rho}_{\rm SNII}$ measurements provide a strong lower bound on the
normalisation of the SFH. Particularly given that uncertainty regarding
obscuration corrections is more likely to raise than lower the
$\dot{\rho}_{\rm SNII}$ measurements, the SFH normalisation cannot
realistically be much lower than that obtained from assuming the BG IMF
(Figure~\ref{fig:sfhfit}b). The scaling factors for the SFH fits suggest that
this provides further evidence against the $T=8\,$MeV assumption, for
both assumed IMFs. Moreover, the $\dot{\rho}_{\rm SNII}$ measurements are
unlikely to be affected by sufficient obscuration to support an SFH
normalisation much higher than that obtained with the SalA IMF. The level of
obscuration to be inferred in order to make the $\dot{\rho}_{\rm SNII}$
measurements lie above the upper edge of the SalA IMF envelope requires
$E(B-V)\approx0.3$ at $z=0.3$ and $E(B-V)\approx0.5$ at $z=0.7$. These
values are quite extreme, corresponding to correction factors of approximately
$3.3$ and $6$ respectively, and values this high are not even inferred from
the luminosity dependent obscuration corrections of \citet{Hop:04}
for UV data at similar redshifts. Higher SFH normalisations may still be
possible, though, and here we return to the assumptions regarding the
mass range over which we assume stars become core collapse SNe. If we allow
only a mass range of $10<M<30\,M_{\odot}$, the $\dot{\rho}_{\rm SNII}$
predictions from the SalA IMF would be lowered by as much as a factor of
$0.6$ ($-0.2\,$dex). Assuming this mass range would place predictions from
even the traditional Salpeter IMF at such a level (i.e., $\approx -0.2\,$dex
from the confidence region in Figure~\ref{fig:snrate1}a). So, at the expense
of an increased lower limit of integration for core collapse SN production,
IMFs providing quite high normalisations for the SFH can still be made
consistent with the $\dot{\rho}_{\rm SNII}$ measurements.

Another point to be considered here is the issue of a possible neutrino
flux generated from a population of massive stellar objects not observable
as SNII \citep{BBM:01,Heg:03,Str:05,Dai:05}. For example, if stars in the
mass range $30<M<50\,M_{\odot}$ produce the same kind of burst of neutrinos
at the end of their life as SNII, but do not become core collapse SNe,
instead progressing directly to a black hole or other exotic end, then the
neutrino flux inferred from the SFH via an assumed SNII rate will be biased,
and the observed SNII rate density could legitimately be $\approx 10\%$ lower
than that predicted from the SFH using this technique. If even lower mass
progenitors become failed supernovae, then this difference could be even
larger. Certainly more robust information regarding the connection between
the final stages of stellar evolution and neutrino emission would help to
refine this type of analysis, and strengthen the implications regarding the
favoured SFH normalisation and corresponding IMF.

Our general conclusions here are (1) that the IMF cannot be much more shallow
at the high mass end than the BG IMF without predicting values for
$\dot{\rho}_{\rm SNII}$ that are too low \citep[see also][]{Loe:06};
and (2) in the absence of a lower
integration limit as high as $\approx 10\,M_{\odot}$ for core collapse SN
production, the IMF cannot produce SFH normalisations much higher than
does the SalA IMF without predicting values for $\dot{\rho}_{\rm SNII}$
that begin to require quite extreme obscuration corrections. Some slightly
less extreme IMFs, while still producing higher SFH normalisations than
the SalA IMF, may be allowed if the core collapse SN lower mass cut-off
lay between 8 and 10\,$M_{\odot}$. This would still seem to be hard to justify,
though, as evidence is growing for SN progenitors at these low masses. The SNII
analysed by \citet{Van:06}, for example, (2003gd), favours a progenitor mass
$8-9\,M_{\odot}$ \citep[see also][]{Sma:04}. A second SNII (2005cs) also
favors a low progenitor
mass, $9^{+3}_{-2}\,M_{\odot}$ \citep{MSD:05}. To confirm and refine these
IMF constraints, a larger selection of independent $\dot{\rho}_{\rm SNII}$
measurements, spanning a broad range of redshift, would be invaluable.
As noted in \citet{Str:05}, sufficiently precise SNII rate data could be used
to directly predict the DSNB flux, independently of assumptions about the
SFH and IMF.

\subsection{Supernovae type~Ia}

\begin{figure*}
\centerline{\rotatebox{-90}{\includegraphics[width=7cm]{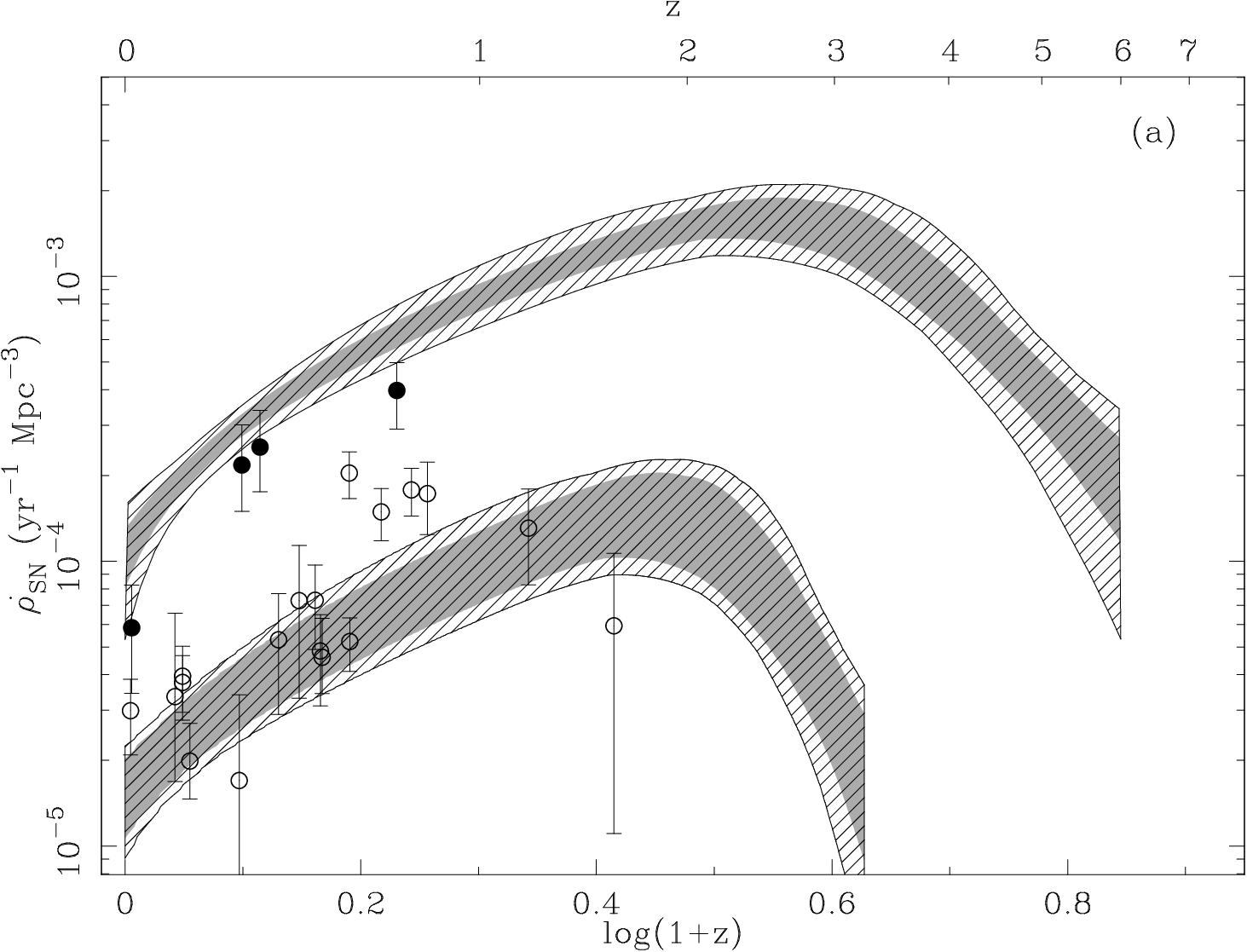}}
\rotatebox{-90}{\includegraphics[width=7cm]{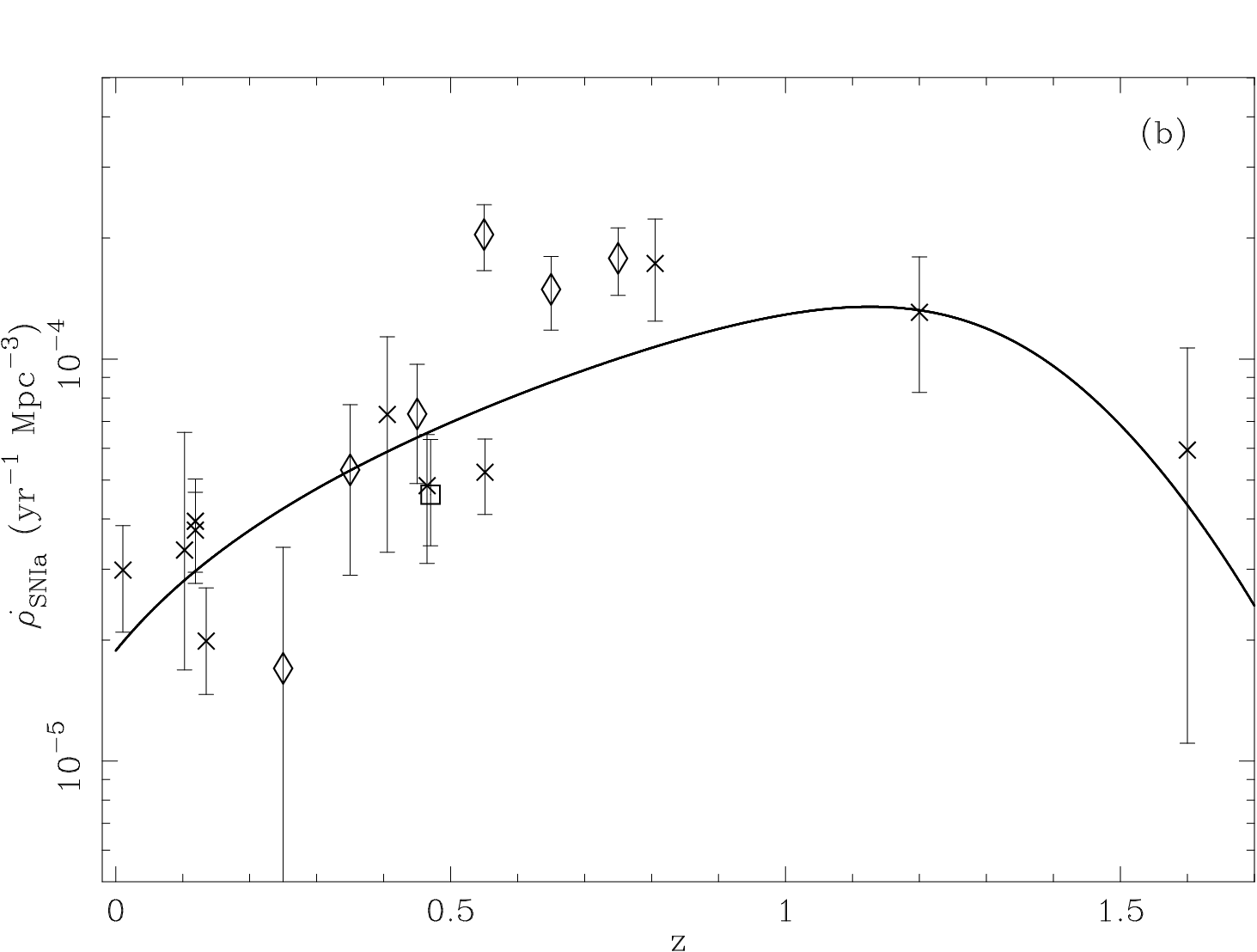}}}
\caption{(a) As for Figure~\ref{fig:snrate1}b, but assuming $t_{Ia}=1\,$Gyr.
(b) The best fit to the SNIa rate from the SFH, with $A=1.15\times10^{-3}$ and
$t_{Ia}=2.7\,$Gyr (see text). Crosses are the data compilation
from \citet{SR:06}, diamonds are data from \citet{Bar:06} and the square
is from \citet{Nei:06}. Note different axes ranges from (a), and the linear
redshift scale.
 \label{fig:snrate2}}
\end{figure*}

The prediction for $\dot{\rho}_{\rm SNIa}$ from the SFH is also particularly
intriguing. The assumption of the fixed $t_{Ia}=3\,$Gyr has the effect of
matching the $z\gtrsim 3$ turnover in the fitted SFH with the apparent
decline in $\dot{\rho}_{\rm SNIa}$ seen in the highest redshift measurement
from the GOODS sample of \citet{Dah:04}. It is possible, indeed probable,
that this is simply a coincidence as it is a single $\dot{\rho}_{\rm SNIa}$
measurement, with large uncertainties, that is suggestive of the decline, and
the turnover in the SFH is driven almost entirely by the $z\approx 6$
measurement of \citet{Bun:04}. It is thus still highly possible that the
decline in both the SFH and $\dot{\rho}_{\rm SNIa}$ lie at somewhat higher
redshift. In particular, recent spectroscopic results from a complete magnitude
limited sample \citep{LeFev:05} suggest that the SFH inferred at $z=3-4$ is up
to two or three times higher than that estimated from colour-selected LBGs
\citep{Ste:99}. This may imply that the shape of the SFH is flatter
between $2<z<6$ than our current fits suggest. Even more tantalisingly,
the higher SFH estimates from the spectroscopic measurements compared to
the colour-selected samples at $3<z<4$ suggest that the photometric dropout
selected samples at even higher redshift ($z\approx 6$) may also be
underestimating the total SFH. This effect would be compounded by
the recent evidence suggesting non-negligible obscuration at high-$z$
\citep{Cha:05,And:05}, and appears to provide evidence that the expected
high redshift decline in the SFH may still not be well established yet.
With this in mind, it is nonetheless interesting to note that
the robustness of the GOODS measurements, again, has been explored in
some detail by \citet{Dah:04} who are confident of the reliability of
this feature in the $\dot{\rho}_{\rm SNIa}$ data. If the turnover in the
SFH also turns out to be reliable, the $\dot{\rho}_{\rm SNIa}$ evolution can
be used to constrain the delay time for SNIa \citep[e.g.,][]{For:06}.
The predictions for $\dot{\rho}_{\rm SNIa}$ are quite different for
$t_{Ia}=1\,$Gyr (Figure~\ref{fig:snrate2}a) compared with $t_{Ia}=3\,$Gyr,
and refining the measurements of $\dot{\rho}_{\rm SNIa}$ between $1<z<3$ would
be quite revealing. Alternatively, if the SFH turnover does indeed lie at $z>6$
that would imply either that $t_{Ia}>3\,$Gyr, a result that appears to lie
outside the currently favoured range, or that the true turnover in the
$\dot{\rho}_{\rm SNIa}$ measurements is also at somewhat higher redshift
than the $z\approx 1.5$ from \citet{Dah:04}.

To explore a more physically motivated connection between SNIa generation
and the underlying stellar populations, \citet{SB:05} introduced a
two component model, dependent on both SFR and stellar mass densities,
to derive the SNIa rate density. This has the effect of allowing a
contribution to the SNIa rate from old stellar populations where the
current SFR may be low, while maintaining a contribution from the
currently star forming systems. Taking a logical next step, \citet{Nei:06}
allow a delay time to be incorporated into such a two component model,
and find a characteristic delay time $\tau=3\,$Gyr (given a
Gaussian distribution of delay times), consistent with our simple
result above, although their model also includes a contribution from a
component proportional to $\rho_*$. \citet{Man:06} find a bimodal distribution
of delay times, with a ``prompt" component having a delay time of
$\approx0.1\,$Gyr. (\citeauthor{Pin:06} \citeyear{Pin:06} find evidence that
the joint IMF of binary stars favors ``twins" of nearly equal mass, and
suggest that this may provide a natural explanation for these prompt SNIa.)
To illustrate the power of a well constrained SFH, we use our best fitting
SFH for the BG IMF (from Table~\ref{tab1}) to explore the available
parameters in a \citet{SB:05} model and, keeping in mind the cautions
regarding the high redshift declines in both $\dot{\rho}_{\rm SNIa}$ and
$\dot{\rho}_*$, do not try to overinterpret our results. We explored the
connection between $\dot{\rho}_{\rm SNIa}$, $\dot{\rho}_*$ and $\rho_*$
using the relation
\begin{equation}
\dot{\rho}_{\rm SNIa}(t) = A\dot{\rho}_*(t-t_{Ia}) + B\rho_*(t)
\end{equation}
with all quantities in the physical units used in this investigation.
Performing a simple $\chi^2$ minimisation, allowing $A$, $B$ and $t_{Ia}$
to vary, we find the best fitting solution favours
$A=1.15\times10^{-3}\,M_{\odot}^{-1}$, $B=0\,M_{\odot}^{-1}$yr$^{-1}$,
and $t_{Ia}=2.7\,$Gyr. This result is illustrated in
Figure~\ref{fig:snrate2}b, and is clearly driven fairly strongly
(as expected intuitively from Figure~\ref{fig:snrate1}) by the
combination of the high-$z$ $\dot{\rho}_{\rm SNIa}$ measurement and
the turnover in the SFH driven by the \citet{Bun:04} measurement.
The value of $A=1.15\times10^{-3}\,M_{\odot}^{-1}$ is consistent with the
range of $0.001 \le f_{Ia} \le 0.0014 \,M_{\odot}^{-1}$ assumed for the limits
in Figure~\ref{fig:snrate1}. The $B=0$ result seems to arise from the strong
similarity in shape between the SFH and the $\dot{\rho}_{\rm SNIa}(z)$
evolution, while the $\rho_*(z)$ evolution has the opposite shape (increasing
to lower redshifts, rather than decreasing). It seems clear that allowing
negative values of B would allow some non-zero solution, although
what physical interpretation this would have is not clear. It certainly
would not be in the spirit of the model as intended by \citet{SB:05}.
As concluded above, and since $B=0$, our tentative inference appears to
favour SNIa delay times close to $t_{Ia}\approx 3\,$Gyr, although there
are clearly much more sophisticated models, including those allowing
for a distribution in delay times, that we have not explored here,
and the locations of the apparent downturns in both $\dot{\rho}_{\rm SNIa}$
and $\dot{\rho}_*$ clearly have a very strong influence on this result.

\subsection{Detecting the DSNB}

With the best fitting SFH models explored here, the predictions for
the DSNB appear to lie excitingly close to the measured $\overline{\nu}_e$
flux limit (Figure~\ref{fig:sneas}). It is clear that directly observing the
DSNB will allow much greater insight into the properties of star formation.
Already the DSNB constraint indicates a preferred IMF range and normalisation
for the SFH. It also illustrates that stronger constraints on the SFH have
implications for understanding the details of both SNII and SNIa production,
and the physical basis of neutrino generation by SNII is intimately associated
with all these predictions. Being able to detect the DSNB and its energy
spectrum will allow a more sophisticated analysis of the detailed connections
between all these aspects of star formation and the cosmic SFH.

Methods for increasing the sensitivity of particle detectors to DSNB
antineutrinos and neutrinos have been detailed elsewhere, and we briefly
reiterate some of these conclusions, to illustrate the potential for detecting
the DSNB. \citet{BV:04} propose loading SK with dissolved gadolinium
trichloride to allow tagging of neutron captures, thus significantly lowering
backgrounds, in order to directly detect the DSNB $\overline{\nu}_e$ spectrum.
This provides perhaps the best immediate possibility for measuring the DSNB
spectrum shape, which, given the minimum normalisation of the SFH assuming the
BG IMF, seems to be lying just below the current flux limit.
\citet{BS:06} describe how the Sudbury Neutrino Observatory (SNO) could
improve the current flux limit for DSNB $\nu_e$ by about three orders
of magnitude by coupling the background analysis from SK with the sensitivity
to $\nu_e$ at SNO. Combining information on both DSNB $\nu_e$ and
$\overline{\nu}_e$ populations will allow further exciting insight and
constraints on SNII neutrino production \citep[the connection between the
two is explored further by][]{Lun:06}.

\section{Summary}
\label{summ}

We have updated the SFH compilation of \citet{Hop:04}, emphasising the
strong constraints from recent UV and FIR measurements, and refining
the results of numerous measurements over the past decade. An analysis of
various uncertainties that may contribute to the normalisation of the SFH
has been explored and the IMF assumptions play a key role, being
essentially the only assumption that can lower the normalisation.
We performed parametric fits to the SFH, using both the form of
\citet{Col:01} and piecewise linear models, both constrained by the
SK $\overline{\nu}_e$ limit. The results suggest that the preferred IMF
should produce SFH normalisations within the range of those from the
modified Salpeter A IMF \citep{Bal:03} and the IMF of \citet{Bal:03}. They
also suggest that lower temperatures ($T=4-6\,$MeV) are preferred for the
$\overline{\nu}_e$ population. It should be noted that here we have
assumed a simple Fermi-Dirac spectrum for the $\overline{\nu}_e$ spectrum
after neutrino mixing. Since the current SK energy threshold ($19.3\,$MeV)
is so high, however, the DSNB flux limit is sensitive to the assumed shape
of the spectrum, and a reduction in the height of the spectral tail would
allow a higher average neutrino energy. This highlights the importance of
lowering the energy threshold \citep{BV:04,Yuk:05}. Additionally, a more
accurate treatment of future data would be based not on the integrated flux
above the energy threshold, but rather on the detected event rate spectrum
above the energy threshold \citep{Yuk:05}, after weighting with the neutrino
interaction cross section \citep{VB:99}.

Based on the fits to the SFH we predict the evolution of $\rho_*$,
$\rho_{Z}$ and $\dot{\rho}_{\rm SN}$, and compare with observations.
The comparisons with $\dot{\rho}_{\rm SN}$ are most revealing, providing
the limitations on the assumed IMF, and tentatively favouring longer delay
times ($t_{Ia}=3\,$Gyr) for the SNIa population. The $\dot{\rho}_{\rm SN II}$
measurements provide a key constraint on the SFH normalisation, and
correspondingly on the favoured IMF. In particular, these data bound the
SFH from {\it below}, while the DSNB bounds the SFH from {\it above}.
(Due to the SK energy threshold, the DSNB flux limit primarily constrains
SNII with $z \lesssim 1$, the same range in which we wish to test the
factors that normalize the SFH.) Together, these provide a novel technique
for testing or verifying measurements of a universal IMF, and emphasises
the importance of understanding the range of stellar masses leading to
the various stellar evolution core collapse outcomes. More measurements of
$\dot{\rho}_{\rm SN}$ for both type~II and Ia SNe over a broader redshift
range \citep{Oda:05,Mes:06} would help to more strongly constrain both the
preferred universal IMF and the properties of SNe. Observing the high
redshift turnover in the SNIa rate would also have strong implications for
the location of the expected high redshift turnover in the SFH. Direct
observation of the DSNB will clearly allow much greater insight into the
physics and astrophysics of star formation and supernovae. Given the power
of the existing SK $\overline{\nu}_e$ flux limit, any improvements in
sensitivity will have a very strong impact on constraining the product of the
dust corrections, IMF normalisation, and neutrino emission per supernova.
In fact, since reasonable choices for all of these already saturate the SK
limit, we expect that the significantly improved sensitivity which would be
enabled by adding gadolinium to SK \citep{BV:04} should lead to a first
detection of DSNB $\overline{\nu}_e$ \citep{BV:04,Str:04,Str:05,Dai:05,Yuk:05}.

\acknowledgements
The authors would like to thank the referee, Shaun Cole, for rectifying
a misconception in an early draft of this manuscript, and for helpful
suggestions leading to the current and much improved version.
We also thank Andy Bunker, Mark Fardal, Louie Strigari and
Mark Sullivan for helpful comments and interesting discussions, and Don Neill
for providing a copy of \citet{Nei:06} prior to submission.
AMH acknowledges support provided by the Australian Research Council
in the form of a QEII Fellowship (DP0557850).
JFB acknowledges support from The Ohio State University and
NSF CAREER grant No.\ PHY-0547102.

\begin{deluxetable}{ccc}
\tablewidth{0pt}
\tablecaption{SFH parametric fitting to the form of \citet{Col:01}.
 \label{tab1}}
\tablehead{
\colhead{parameter} & \colhead{modified Salpeter A IMF\tablenotemark{a}} & \colhead{\citet{Bal:03}\tablenotemark{b}}
}
\startdata
$a$ & 0.0170 & 0.0118 \\
$b$ & 0.13 & 0.08 \\
$c$ & 3.3 & 3.3 \\
$d$ & 5.3 & 5.2
\enddata
\tablenotetext{a}{For this fit $\chi^2=37.5$. The scaling factors assuming
$T=(4,6,8)\,$MeV are $(1.0,1.0,0.67)$.}
\tablenotetext{b}{For this fit $\chi^2=37.5$. The scaling factors assuming
$T=(4,6,8)\,$MeV are $(1.0,1.0,0.74)$.}
\end{deluxetable}

\begin{deluxetable}{ccc}
\tablewidth{0pt}
\tablecaption{Piecewise linear SFH parametric fitting.
 \label{tab:lines}}
\tablehead{
\colhead{parameter\tablenotemark{a}} & \colhead{modified Salpeter A IMF\tablenotemark{b}} & \colhead{\citet{Bal:03}\tablenotemark{c}}
}
\startdata
$a$ & $-1.82$ & $-2.02$ \\
$b$ & $3.28$ & $3.44$ \\
$c$ & $-0.724$ & $-0.930$ \\
$d$ & $-0.26$ & $-0.26$ \\
$e$ & $4.99$ & $4.64$ \\
$f$ & $-8.0$ & $-7.8$ \\
$z_1$ & $1.04$ & $0.97$ \\
$z_2$ & $4.48$ & $4.48$ \\
\enddata
\tablenotetext{a}{The parameters here are the intercepts and slopes of
the linear segments in $\log(1+z)-\log(\dot{\rho}_*)$ space, and the
redshifts at which the slope changes. The parameters $a, b$ are the
intercept and slope for the line segment between $0\le z\le z_1$;
$c, d$ are intercept and slope between $z_1\le z\le z_2$;
$e, f$ are intercept and slope between $z_2\le z\le 6$. All
eight parameters are shown for convenience, but $c$ and $e$ are not
independent and are not free parameters in the fitting.}
\tablenotetext{b}{For this fit $\chi^2=19.8$. The scaling factors assuming
$T=(4,6,8)\,$MeV are $(1.0,1.0,0.63)$.}
\tablenotetext{c}{For this fit $\chi^2=19.6$. The scaling factors assuming
$T=(4,6,8)\,$MeV are $(1.0,1.0,0.67)$.}
\end{deluxetable}

\end{document}